%% file: main.tex
\documentclass[10pt,journal]{IEEEtran}
\usepackage{amsmath,amsfonts}
\usepackage{array}
\usepackage[caption=false,font=normalsize,labelfont=sf,textfont=sf]{subfig}
\usepackage{textcomp}
\usepackage{stfloats}
\usepackage{url}
\usepackage{verbatim}
\usepackage{graphicx}
\usepackage[noadjust]{cite} 

\usepackage{algorithm}      
\usepackage{algpseudocode}  

\PassOptionsToPackage{hyphens}{url} 
\usepackage[hidelinks,breaklinks]{hyperref} 

\usepackage{balance}
\usepackage{multirow}

\usepackage[all=normal,mathspacing]{savetrees} 

\newcommand{\reb}[1]{{\color{black}#1}}
\usepackage[export]{adjustbox}

\newcommand{\1}{\mathbf{1}}
\newcommand{\e}{\mathbf{e}}
\newcommand{\0}{\mathbf{0}}
\newcommand{\I}{\mathbf{I}}
\newcommand{\abf}{\mathbf{a}}

\DeclareMathOperator*{\argmax}{argmax}
\DeclareMathOperator*{\argmin}{argmin}

\newcommand{\x}{\mathbf{x}}
\newcommand{\s}{\mathbf{s}}
\newcommand{\cbf}{\mathbf{c}}
\newcommand{\y}{\mathbf{y}}

\newcommand{\D}{\mathbf{D}}
\newcommand{\G}{\mathbf{G}}

\newcommand{\vbf}{\mathbf{v}}
\newcommand{\U}{\mathbf{U}}
\newcommand{\V}{\mathbf{V}}
\newcommand{\W}{\mathbf{W}}

\newcommand{\Lambd}{\boldsymbol{\Lambda}}

\newcommand{\Q}{\mathbf{Q}}
\newcommand{\R}{\mathbf{R}}
\newcommand{\Mbf}{\mathbf{M}}
\newcommand{\Sbf}{\mathbf{S}}
\newcommand{\pbf}{\mathbf{p}}
\newcommand{\Pbf}{\mathbf{P}}
\newcommand{\Real}{\mathbb{R}}
\newcommand{\dbf}{\mathbf{d}}
\newcommand{\z}{\mathbf{z}}

\newcommand{\m}{\mathbf{m}}

\newcommand{\taubf}{\boldsymbol{\tau}}

\newcommand{\Sigmabf}{\boldsymbol{\Sigma}}
\newcommand{\ji}{\jmath}

\DeclareMathOperator{\diag}{\rm{diag}}
\DeclareMathOperator{\EX}{\mathbb{E}}
\usepackage{tikz,pgfplots}
\usetikzlibrary{decorations.pathreplacing,plotmarks}
\pgfplotsset{compat=1.7}
\usepackage{textcomp}
\usetikzlibrary{shapes,arrows}
\usetikzlibrary{arrows.meta}
\usepackage{mathtools}
\usepackage{environ}
\makeatletter
\tikzset{%
  block/.style    = {draw, thick, rectangle, minimum height = 3em,
    minimum width = 3em},
  sum/.style      = {draw, circle, node distance = 2cm}, 
    input/.style    = {coordinate}, 
  output/.style   = {coordinate} 
}
\newsavebox{\measure@tikzpicture}
\NewEnviron{scaletikzpicturetowidth}[1]{%
  \def\tikz@width{#1}%
  \begin{lrbox}{\measure@tikzpicture}%
  \BODY
  \end{lrbox}%
  \pgfmathparse{#1/\wd\measure@tikzpicture}%
  \BODY
}
\usetikzlibrary{calc}
\def\centerarc[#1](#2)(#3:#4:#5)
    { \draw[#1] ($(#2)+({#5*cos(#3)},{#5*sin(#3)})$) arc (#3:#4:#5); }

\tikzset{
 thicker/.style={line width=#1\pgflinewidth},
 thicker/.default={2},
}
\def\mic(#1)(#2)(#3);{
    \begin{scope}[shift={(#1)},rotate={#2},scale={#3}]
        \draw (0,0) circle (0.22);
        \draw[thicker=1.5] 
            (-0.22,-0.3) -- (-0.22,0.3);
    \end{scope}
}
\usetikzlibrary{shapes.geometric}
\def\speaker(#1)(#2)(#3);{
    \begin{scope}[shift={(#1)},rotate={#2},scale={#3}]
        \draw (-0.8,-0.40) rectangle (-0.35,0.40);
        \draw[-] (-0.35,0.40) -- (0,0.65) -- (0,-0.65) -- (-0.35,-0.40);
    \end{scope}
}

\newenvironment{lbmatrix}[1]
  {\left[\array{@{}*{#1}{c}@{}}}
  {\endarray\right]}

\usepackage{empheq}

\usepackage{tabularx}

\begin{document}

\title{Multi-Source Position and Direction-of-Arrival Estimation Based on Euclidean Distance Matrices}

\author{
Klaus Br{\"{u}}mann,~\IEEEmembership{Graduate Student Member,~IEEE,} and Simon Doclo,~\IEEEmembership{Senior Member,~IEEE} \thanks{The authors are with the Department of Medical Physics and Acoustics and the Cluster of Excellence Hearing4all, Carl von Ossietzky Universit\"{a}t Oldenburg, Germany (e-mail: klaus.bruemann@uol.de, simon.doclo@uol.de). This work was funded by the Deutsche Forschungsgemeinschaft (DFG, German Research Foundation) under Germany's Excellence Strategy -- EXC 2177/2 - Project ID 390895286 and Project ID 352015383 - SFB 1330 B2.
} 
}


\maketitle

\begin{abstract}
A popular method to estimate the positions or directions-of-arrival (DOAs) of multiple sound sources using an array of microphones is based on steered-response power (SRP) beamforming. For a three-dimensional scenario, SRP-based methods need to jointly optimize three continuous variables for position estimation or two continuous variables for DOA estimation. This can be computationally expensive, especially when high localization accuracy is desired. In this paper, we propose novel methods for multi-source position and DOA estimation by exploiting properties of Euclidean distance matrices (EDMs) and their respective Gram matrices. All methods require estimated time-differences of arrival (TDOAs) between the microphones. In the proposed multi-source position estimation method only a single continuous variable, representing the distance between each source and a reference microphone, needs to be optimized. For each source, the optimal continuous distance variable and set of candidate TDOA estimates are determined by minimizing a cost function that is defined using the eigenvalues of the Gram matrix. The estimated relative source positions are then mapped to estimated absolute source positions by solving an orthogonal Procrustes problem for each source. The proposed multi-source DOA estimation method entirely eliminates the need for continuous variable optimization by defining a relative coordinate system per source such that one of its coordinate axes is aligned with the respective source DOA. The optimal set of candidate TDOA estimates is determined by minimizing a cost function that is defined using the eigenvalues of a rank-reduced Gram matrix. 
For two sources in a noisy and reverberant environment, experimental results for different source and microphone configurations with six microphones show that the proposed EDM-based method consistently outperforms the SRP-based method in terms of position and DOA estimation accuracy. 
{Despite the fact that the computational complexities of the EDM-based methods grow combinatorially with the number of microphones and candidate TDOA estimates, they exhibit lower run times than the SRP-based methods for the considered experimental setup with a moderate number of microphones.}
\end{abstract}

\begin{IEEEkeywords}
Source localization, position estimation, direction-of-arrival estimation, multi-source, Euclidean distance matrix, Gram matrix, rank, time-difference of arrival.
\end{IEEEkeywords}

\input{Sec_Introduction.tex}

\input{Sec_Theory.tex}

\input{Sec_EDM_based_Position_Estimation.tex}

\input{Sec_EDM_based_DOA_Estimation.tex}

\input{Sec_SRP_PHAT.tex}

\input{Sec_Experimental_Evaluation.tex}

\section{Conclusions}\label{sec: Conclusions}
In this paper, we have proposed novel TDOA-based multi-source position and DOA estimation methods that exploit properties of Euclidean distance matrices. 
For 3D position estimation, the proposed method requires optimizing only a single continuous variable instead of jointly optimizing three position variables as in SRP-based methods. 
This variable, representing the distance between each source and a reference microphone, is optimized for each combination of candidate TDOA estimates with a cost function based on the eigenvalues of the Gram matrix associated with the EDM. 
The estimated relative source positions, obtained from the Gram matrices corresponding to the optimal cost function values, are then mapped to absolute source positions by solving an orthogonal Procrustes problem for each source. 
For 3D DOA estimation, we define a relative coordinate system for each source, with one axis aligned to the DOA vector. 
The optimal set of candidate TDOA estimates is determined by minimizing a cost function based on the eigenvalues of a rank-reduced Gram matrix, requiring no continuous variable optimization. 
The source DOA vectors are then estimated by mapping the relative microphone positions, obtained from the rank-reduced Gram matrices, to the absolute microphone positions, for each source. 
For two sources in a noisy and reverberant environment, experimental results across a wide range of source and microphone configurations, including compact as well as distributed microphone arrays, show that the proposed EDM-based methods consistently outperform the SRP-based methods in terms of position and DOA estimation accuracy. 
{Although the computational complexities of the EDM-based methods grow combinatorially with the number of microphones, due to the reduction of the number of continuous variables to be jointly optimized, the run times of the EDM-based methods are lower than the SRP-based methods for the considered experimental setups with six microphones.} 
In future work, it would be interesting to \reb{explore the influence of stronger reverberation on the practical choice of the number of candidate TDOA estimates and to} combine the estimated 3D positions or DOAs from the EDM-based methods with source tracking methods \cite{Ward2003ParticleFiltering, Gannot2006MicrophoneArrayLocalizers, Evers2020LOCATA} for moving source scenarios.

\bibliographystyle{IEEEtran}
\bibliography{ms}

\newpage

 

\vspace{11pt}
\begin{IEEEbiography}[{\includegraphics[width=1in,height=1.25in,clip,keepaspectratio]{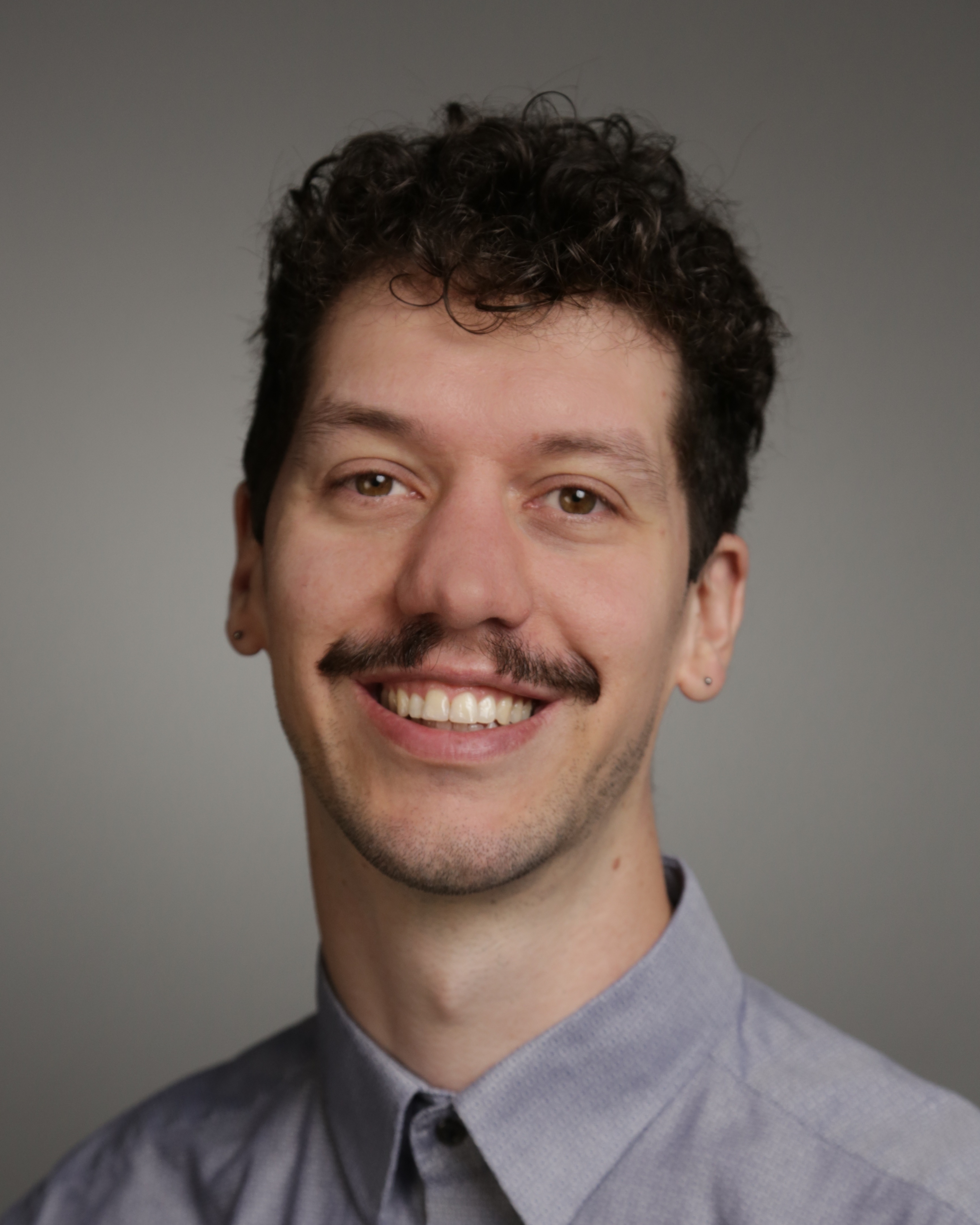}}]{Klaus Br\"{u}mann}
(Student Member, IEEE) received the B.Eng. and M.Sc. degrees in Engineering
Physics from Carl von Ossietzky Universit\"{a}t Oldenburg, Germany, in 2019 and 2021,
respectively. Since 2021, he has been working towards obtaining
the Dr.-Ing. degree with the Signal Processing Group, 
Carl von Ossietzky Universit\"{a}t Oldenburg, Germany, focusing
on the analysis and development of multi-source localization methods using compact and distributed microphones. 
\end{IEEEbiography}

\vspace{11pt}
\begin{IEEEbiography}[{\includegraphics[width=1in,height=1.25in,clip,keepaspectratio]{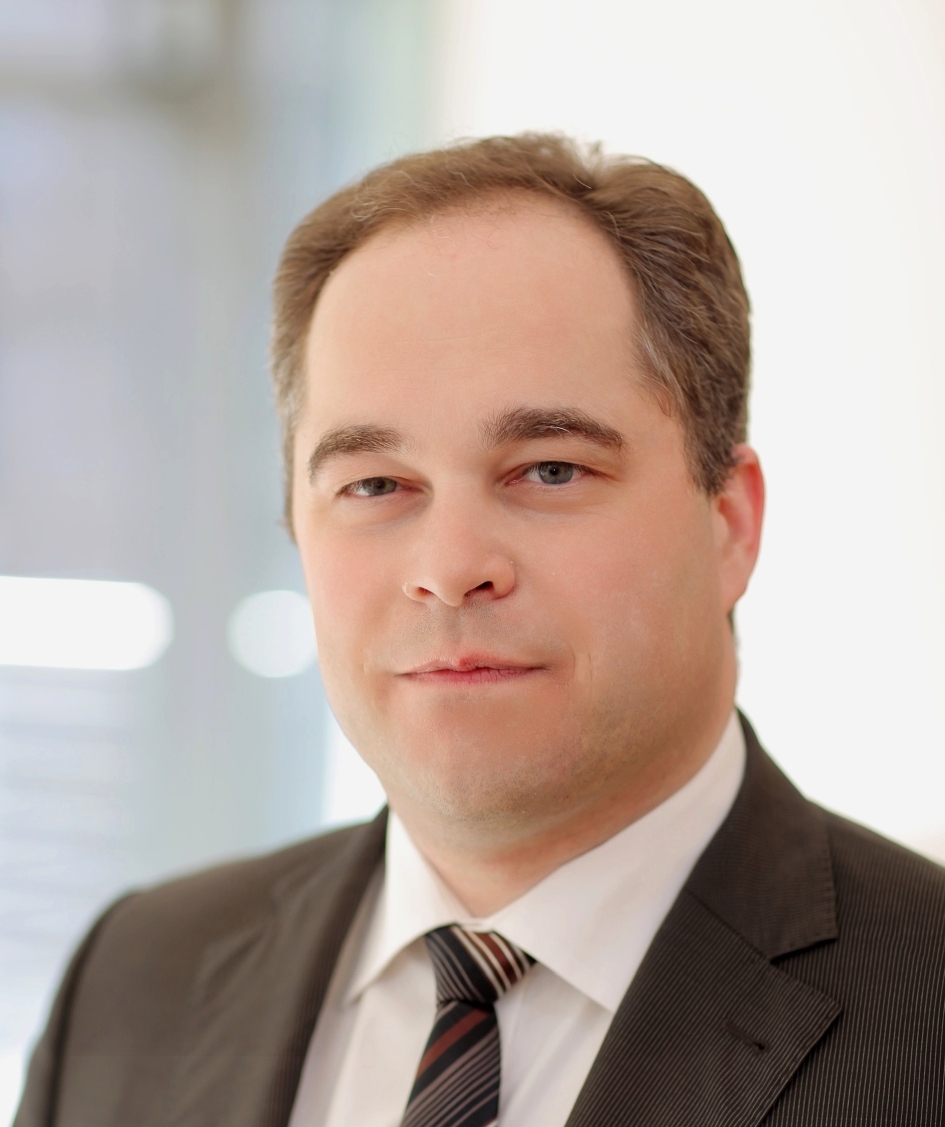}}]{Simon Doclo}
(S'95-M'03-SM'13) received the M.Sc. degree in electrical engineering 
and the Ph.D. degree in applied sciences from KU Leuven, Belgium, in 1997 and 2003. From 2003 to 2007 he was 
a Postdoctoral Fellow at KU Leuven and McMaster University, Canada. From 2007 
to 2009 he was a Principal Scientist with NXP Semiconductors in Leuven, Belgium. 
Since 2009 he is a full professor at the University of Oldenburg, Germany, and scientific advisor for the 
Branch Hearing, Speech and Audio Technology HSA of the Fraunhofer Institute for Digital Media Technology IDMT. 
His research activities center around signal processing for acoustical and biomedical applications, 
more specifically microphone array processing, speech enhancement, active noise control, 
acoustic sensor networks and hearing aid processing.
Prof. Doclo received several best paper awards (International Workshop on Acoustic Echo and Noise Control 2001, EURASIP Signal 
Processing 2003, IEEE Signal Processing Society 2008, VDE Information Technology Society 2019, ITG Conference on Speech Communication 2025). He served as associate editor for EURASIP Journal on Advances in Signal Processing (2014-2019) and senior area editor for IEEE Transactions on Audio, Speech and Language Processing (2021-2025). He was member of the IEEE Signal Processing Society Technical Committee on Audio and Acoustic Signal Processing and the EURASIP Technical Area Committee on Acoustic, Speech and Music Signal Processing, and is member of the EAA Technical Committee on Audio Signal Processing. 
\end{IEEEbiography}

\vfill
\end{document}

%% file: Sec_Introduction.tex
\section{Introduction}
\label{sec: Introduction}
\IEEEPARstart{E}{stimating} the locations, i.e., positions or directions-of-arrival (DOAs), of multiple speech sources in a noisy and reverberant environment is important for several applications, such as source tracking, 
speech enhancement, and speaker extraction \cite{madhu2008acoustic, pertila2018multichannel, aroudi2020cognitive, tesch2023multi}. 
In this paper, we consider both compact microphone arrays with relatively small inter-microphone distances as well as spatially distributed microphones with large inter-microphone distances. 
For compact microphone arrays the sources will be assumed to be in the far field, such that we will focus on estimating the DOA (azimuth and elevation) of each source, whereas for spatially distributed microphones we will focus on estimating the three-dimensional position of each source. 


Existing model-based methods for source localization fall into two main categories \cite{madhu2008acoustic, pertila2018multichannel}: one-step methods such as steered-response power methods \cite{omologo1997use, dibiase2001robust, brutti2010multiple} or the subspace-based multiple signal classification (MUSIC) method \cite{schmidt1986multiple}, and two-step methods, which rely on the prior estimation of variables such as time-differences of arrival (TDOAs) between microphones \cite{huang2008time}. 
In addition, several learning-based methods \cite{yalta2017sound, adavanne2021differentiable, bianco2021semi, grumiaux2022survey, yang2022srp, grinstein2024neural, varzandeh2025improving} for source localization have recently been proposed, showing promising results. 
However, most learning-based methods are trained for a specific array geometry and do not support source localization with arbitrary array geometries. 
In this paper, we only consider model-based localization methods, which provide the flexibility to use different array geometries without the need to retrain a neural network. 
A popular model-based method for single-source DOA or position estimation is based on the steered-response power with phase transform (SRP-PHAT) functional \cite{dibiase2001robust, cobos2010modified, nunes2014steered, salvati2017exploiting, long2018acoustic, garcia2021analytical, grinstein2024steered, dietzen2024scalable}. 
While DOA estimation requires jointly optimizing the SRP-PHAT functional in up to two continuous variables (azimuth and elevation), position estimation requires jointly optimizing the SRP-PHAT functional in up to three continuous variables (x, y, and z coordinates). 
To perform accurate source localization, the SRP-PHAT functional needs to be evaluated on a discrete multi-dimensional grid with a high enough resolution in each variable resulting in a high computational complexity \cite{garcia2021analytical, huang2021robust}. 
To avoid the multi-dimensional search of SRP-based methods, in \cite{brumann20223d} we proposed a method to estimate the three-dimensional position of a single source based on Euclidean distance matrices. 
This method minimizes a cost function defined using the eigenvalues of a Gram matrix associated with an EDM containing the inter-microphone distances and the source-microphone distances. 
Using estimated TDOAs, this problem can be reformulated as an optimization in only a single distance variable, where the solution represents the distance between the source and the reference microphone. 
Several model-based methods have also been proposed\linebreak{} for multi-source DOA and position estimation, e.g., based on determining multiple peaks in the SRP-PHAT functional or MUSIC spectrum \cite{schmidt1986multiple, brutti2010multiple, oualil2013fast, salvati2014incident, gerlach2014joint, fejgin2024comparison} often using clustering/association of acoustic features \cite{oualil2013fast, liu2021multiple, dang2021feature}. 
It should be noted that most of these methods focus on multi-source DOA estimation and not on multi-source position estimation. 

In this paper, we first extend the EDM-based single-source position estimation method in \cite{brumann20223d} to multi-source position estimation, assuming the number of sources $S$ to be known. 
For each combination of candidate TDOA estimates, the optimal distance variable is determined by minimizing the EDM-based cost function using a one-dimensional exhaustive search. 
Assuming that each source position corresponds to a unique combination of TDOAs\footnote{\color{black}The probability that sources have the same TDOAs in all pairs of microphones is extremely low. This would only occur for a planar (or linear) microphone array with symmetric sources with respect to this plane (or line). In this paper, we focus on three-dimensional microphone configurations.\color{black}}, the optimal set of candidate TDOA estimates and corresponding source distances are determined, thereby also solving the association of candidate TDOA estimates to sources. 
The estimated relative position of each source to the microphone array can then be determined from the Gram matrix with the estimated source distance. 
The estimated relative positions are then mapped to estimated absolute source positions by solving an orthogonal Procrustes problem for each source. 
As the second contribution of this paper, we propose an EDM-based method to estimate the DOAs of multiple sources relative to a compact microphone array, which does not require continuous variable optimization. 
We construct a rank-reduced Gram matrix by subtracting a rank-1 matrix based on the estimated TDOAs from the Gram matrix associated with the EDM containing the microphone distances. 
We define a cost function using the eigenvalues of this rank-reduced Gram matrix and, similarly as for multi-source position estimation, determine the optimal set of candidate TDOA estimates by considering the $S$ smallest cost function values. 
The DOA of each source is then estimated from the mapping between the absolute coordinate system and the relative coordinate system, which depends on the combinations of TDOA estimates corresponding to each source. 

Two sets of experiments are conducted for several simulated scenarios with two sources and six microphones in a room with mild background noise and reverberation. 
The results of the first experiment considering spatially distributed microphones show that the proposed EDM-based method outperforms the SRP-based method in estimating the positions of the sources for all considered distances between the sources and the microphone array. 
Similarly, the results of the second experiment considering compact microphone arrays show that the proposed EDM-based method also consistently outperforms the SRP-based method in estimating the DOAs of the sources. 
In terms of computational complexity, the complexities of the EDM-based methods grow combinatorially with the number of microphones and the number of candidate TDOA estimates, whereas the complexities of SRP-based methods only grow quadratically in terms of the number of microphones. 
At the same time, the EDM-based methods reduce the number of continuous variables that need to be jointly optimized, resulting in lower run times. 
\reb{The experimental results also demonstrate that position estimation performance can be improved by setting the number of candidate TDOA estimates equal to the number of sources plus one.}

This paper is organized as follows. 
In Section \ref{sec: Theory}, we introduce the acoustic scenario and the theoretical background and properties of EDMs. 
After reviewing the EDM-based position estimation method for a single source, we extend this method for multi-source position estimation in Section \ref{sec: EDM-Based Position Estimation}. 
In Section \ref{sec: EDM DOA Estimation} we propose EDM-based methods for single-source and multi-source DOA estimation considering far-field sources and compact microphone arrays. 
After discussing the baseline SRP-based method in Section \ref{sec: SRP-PHAT}, we compare the performance of the EDM-based and SRP-based methods for multi-source position and DOA estimation in Section \ref{sec: Evaluation}. 

%% file: Sec_Theory.tex
\section{Theoretical Background}\label{sec: Theory}

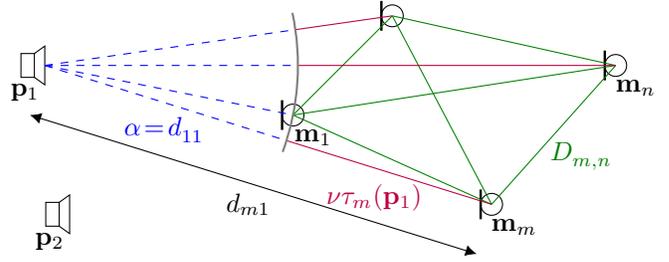
\begin{figure}
\centering
\begin{tikzpicture}[scale = 0.66]
\speaker(12,-11)(0)(0.6);
\node[align=center] at (12-0.4,-11-0.55) {$\pbf_1$};

\speaker(12.5,-14)(0)(0.6);
\node[align=center] at (12.5-0.4,-14-0.55) {$\pbf_2$};

\mic(17,-12)(0)(1);
\node[align=center] at (17+0.4,-12-0.45) {$\m_1$};
\mic(23.5,-11)(0)(1);
\node[align=center] at (23.5+0.4,-11-0.45) {$\m_n$};
\mic(21,-13.8)(0)(1);
\node[align=center] at (21+0.45,-13.8-0.45) {$\m_m$};
\mic(11.6+0.4+18-11,3.6+0.4-14)(0)(1);

\draw[green!50!black,-] (10+18-11,2-14) -- (16.5+18-11,-11); 
\draw[green!50!black,-] (10+18-11,2-14) -- (14+18-11,-13.8); 
\draw[green!50!black,-] (10+18-11,2-14) -- (12+18-11,-10); 
\draw[green!50!black,-] (16.5+18-11,3-14) -- (12+18-11,-10); 
\draw[green!50!black,-] (14+18-11,0.2-14) -- (12+18-11,-10); 
\draw[green!50!black,-] (16.5+18-11,3-14) -- (14+18-11,-13.8); 
\node[align=center] at (22.25+0.55,-12.525-0.3) {$\color{green!50!black}D_{m,n}^{}\color{black}$};

\draw[blue,dashed] (12,-11) -- (17,-12); 
\draw[blue,dashed] (12,-11) -- (17.0990,-11); 
\draw[blue,dashed] (12,-11) -- (17.0478,-10.2789); 
\draw[blue,dashed] (12,-11) -- (16.8688,-12.5147); 
\node[align=center] at (14.4,-12.3) {$\color{blue}\alpha = d_{11}^{}\color{black}$};

\draw[purple,-] (17.0990,-11) -- (23.5, -11); 
\draw[purple,-] (17.0478,-10.2789) -- (19, -10); 
\draw[purple,-] (16.8688,-12.5147) -- (21, -13.8); 
\node[align=center] at (18.9344-0.3,-13.1574-0.5) {$\color{purple} \nu \tau_{m}(\pbf_{1})\color{black}$};

\draw[black,{Latex[length=1.5mm, width=1.5mm]}-{Latex[length=1.5mm, width=1.5mm]}] (12-0.3,-11-1) -- (21-0.3, -13.8-1); 
\node[align=center] at (16.3-0.2,-13.1-0.7) {$d_{m1}$};

\centerarc[gray,thick](12,-11)(-20:12:5.099)
\end{tikzpicture}
\caption{Exemplary acoustic scenario with two sources at positions $\pbf_1$ and $\pbf_{2}$ and a microphone array at positions $\m_{1}, \dots, \m_{M}$, with the first microphone defined as the reference microphone.\vspace*{-0.2 cm}} \vspace*{-4 mm} 
\label{fig: Near field}
\end{figure}

We consider a noisy and reverberant acoustic environment with $S$ static sources, where $\pbf_s \in \Real^P$ denotes the position of the $s$-th source in the absolute coordinate system with $P$-dimensional canonical basis vectors $\e_{1}^{}, \dots, \e_{P}^{}$ ($1 \leq P \leq 3$). 
We assume the number of sources $S$ to be known. 
We consider an array with $M$ microphones whose positions are assumed to be known, with $M > P$, where $\m_{m} \in \Real^P$ denotes the position of the $m$-th microphone. 
The $P \times M$-dimensional microphone positions matrix \color{black} spans $P$-dimensional space and \color{black} is defined as $\Mbf \;=\; [\m_{1}, \dots, \m_{M}]$.
The origin of the coordinate system is defined at the centroid of the microphone positions, i.e., $\frac{1}{M}\Mbf\1_{M}=\0_{P}$,\linebreak{} with $\1_{M}$ denoting an $M$-dimensional vector of ones and $\0_{P}$ denoting a $P$-dimensional vector of zeros. 
The source and microphone positions are exemplified in Fig. \ref{fig: Near field}. 

The $P\times (M+1)$-dimensional microphones and source positions matrix for the $s$-th source is defined as $\Pbf_{s}=[\Mbf ,\; \pbf_{s}]$. %
The $M \times M$-dimensional Gram matrix of the microphones $\G_{MM}$ and the $(M+1) \times (M+1)$-dimensional Gram matrix for the $s$-th source $\G_{s}$ are defined as\newline{}
\noindent
\begin{minipage}{0.48\linewidth}
$$
\G_{MM} \;=\; \Mbf_{}^{\textrm{T}} \Mbf_{}^{} \; , 
$$
\end{minipage}
\hfill
\begin{minipage}{0.48\linewidth}
\begin{equation}
\G_{s} \;=\; \Pbf_{s}^{\textrm{T}} \Pbf_{s}^{} \;,\label{eq: Gram}
\end{equation}
\end{minipage}\vspace{2mm}
where $\{\cdot\}^{\textrm{T}}$ denotes the transpose operator. 
An important property of both Gram matrices, which will be used throughout this paper, is that their rank is at most $P$. 
When considering another orthonormal basis, relative to the original canonical basis (obtained by rotating and/or reflecting the original basis), the relative microphone positions matrix $\Mbf_{\textrm{r}}$ and the relative microphones and source positions matrix for the $s$-th source $\Pbf_{s}$ are given by\newline
\noindent
\begin{minipage}{0.36\linewidth}
$$
\;\;\;\;\;\;\;\;\;\;\;\;\;\;\;\;\Mbf_{\textrm{r}} \;=\; \R^{\textrm{T}} \Mbf \; , \;$$
\end{minipage}
\hfill
\begin{minipage}{0.6\linewidth}
\begin{equation}
\Pbf_{\text{r}s} \;=\; \R^{\textrm{T}} \Pbf_{s} \;=\; [\Mbf_{\textrm{r}} \, , \; \R^{\textrm{T}}\pbf_{s}] \; , \label{eq: Pbf rel}
\end{equation}
\end{minipage}\vspace{2mm}
with $\R$ a $P \times P$-dimensional orthogonal matrix, for which $\R^{\textrm{T}} = \R^{-1}$. 

It is important to realize that the Gram matrices in \eqref{eq: Gram} are unaffected by a basis rotation and/or reflection, i.e.,\newline
\noindent
\begin{minipage}{0.48\linewidth}
$$
\G_{MM} \;=\; \Mbf_{\textrm{r}}^{\textrm{T}} \Mbf_{\textrm{r}}^{} \; , 
$$
\end{minipage}
\hfill
\begin{minipage}{0.48\linewidth}
\begin{equation}
\G_{s} \;=\; \Pbf_{\textrm{r}s}^{\textrm{T}} \Pbf_{\textrm{r}s}^{} \;.\label{eq: Gram rel}
\end{equation}
\end{minipage}\vspace{2mm}

The distance between the $i$-th and the $j$-th microphone is denoted by $D_{ij}=||\m_{i}-\m_{j}||_{2}$ and the distance between the $s$-th source and the $m$-th microphone is denoted by $d_{ms}=||\m_{m}-\pbf_{s}||_{2}$. 
We assume that the microphones are time-synchronized and the acoustic waves of the sources propagate freely towards the microphones (i.e., no objects between the sources and the microphones). 
The TDOA of the direct component of the $s$-th source between the $i$-th and $j$-th microphones is given by
\begin{equation}
    \tau_{ij}(\pbf_{s}) \;\;=\;\; \frac{||\m_{j}-\pbf_s||_2 - ||\m_{i}-\pbf_s||_2}{\nu} \;\;=\;\; \frac{d_{js}-d_{is}}{\nu} \; ,
    \label{eq: TDOA}
\end{equation}
where $\nu$ denotes the speed of sound. 
Without loss of generality, we define the first microphone as the reference microphone. 
From \eqref{eq: TDOA}, it can be directly seen that the distance between the $s$-th source and the $m$-th microphone can be written in terms of the distance between the $s$-th source and the reference microphone as
\begin{equation}
    d_{ms} \;=\; d_{1s} \;+\; \nu \tau_{m}(\pbf_{s}) \; ,
    \label{eq: d_ms}
\end{equation}
\color{black}with $d_{1s}$ denoting the distance between the reference microphone $m=1$ and the $s$-th source\color{black}, and $\tau_{m}(\pbf_{s})$ denoting the TDOA between the $m$-th microphone and the reference microphone. 
The $(M+1)\times (M+1)$-dimensional EDM $\D_{s}$ for the $s$-th source is defined as 
\begin{equation}
{\D}_{s} \; = \; 
\begin{bmatrix}\renewcommand{\arraystretch}{1.3}
\begin{array}{c|c}
  \D_{MM}^{} & \dbf_{s}^{} \\
  \hline
  \dbf_{s}^{\textrm{T}} & 0_{}^{}
\end{array}
\end{bmatrix} \; ,
\label{eq: EDM}
\end{equation}
where the submatrix $\D_{MM}^{} = \left[D^{2}_{ij}\right]$ contains the squared inter-microphone distances and the vector $\dbf_{s} = \left[d^{2}_{1s}, d^{2}_{2s}, \dots, d^{2}_{Ms}\right]^{\textrm{T}}$ contains the squared distances between the $s$-th source and the microphones.  
In \cite{gower1982euclidean, dokmanic2015euclidean}, it was shown that the Gram matrix of the microphones ${\G}_{MM}$ in \eqref{eq: Gram} could be written in terms of the EDM $\D_{MM}^{}$ as 
\begin{equation}
    \G_{MM} \; = \; -\frac{1}{2} \, (\I_{M}-\1_{M}^{}\abf_{M}^{\textrm{T}}) \, {\D_{MM}} \, (\I_{M}^{}-\abf_{M}^{}\1_{M}^{\textrm{T}}) \; ,
    \label{eq: EDM to Gram microphones}
\end{equation}
where $\I_{M}$ denotes the $M \times M$-dimensional identity matrix and $\abf_{M}$ denotes the so-called centering vector. 
To ensure that the centroid of the relative microphone positions $\Mbf_{\textrm{r}}$ is at the origin, we define the centering vector for $\G_{MM}$ as\linebreak{} $\abf_{M} = \frac{1}{M}\1_{M}$. 
This important property will be used in the EDM-based DOA estimation in Section \ref{sec: EDM DOA Estimation}. 
Similarly as in \eqref{eq: EDM to Gram microphones}, the Gram matrix for the $s$-th source $\G_{s}$ can be written in terms of the EDM $\D_{s}$ as
\begin{equation}
    {\G}_{s} \; = \; -\frac{1}{2} \, (\I_{M+1}-\1_{M+1}^{}\abf_{M+1}^{\textrm{T}}) \, {\D}_{s} \, (\I_{M+1}^{}-\abf_{M+1}^{}\1_{M+1}^{\textrm{T}}) \; ,
    \label{eq: EDM to Gram}
\end{equation}
where the centering vector is defined as $\abf_{M+1} = \frac{1}{M}[\1_{M}^{\textrm{T}},0]^{\textrm{T}}$, again to ensure that the centroid of the relative microphone positions is at the origin. 
Let us consider the eigenvalue decomposition 
\begin{equation}
    {\G}^{}_{s} \; = \; \Sbf^{}_{} \Lambd^{}_{} \Sbf^{\textrm{T}}_{} \; ,
    \label{eq: Gram EVD}
\end{equation}
with $\Sbf_{}$ and $\Lambd_{}$ denoting the matrices containing the eigenvectors and eigenvalues, respectively. 
Since the rank of the positive semi-definite matrix $\G_{s}$ is at most $P$, this means that at most $P$ eigenvalues are non-zero, i.e., $\lambda_{1} \geq \dots \geq \lambda_{P} \geq 0$ and $\lambda_{P+1} = \dots = \lambda_{M+1} = 0$. 
Using \eqref{eq: Gram rel} and \eqref{eq: Gram EVD}, the relative microphones and source positions matrix $\Pbf_{\textrm{r}s}$ can hence be written as
\begin{equation}
    \Pbf_{\textrm{r}s} \; = \; \left[\textrm{diag}\left(\sqrt{\lambda_{1}}, \dots ,\sqrt{\lambda_{P}}\right) \, , \; \0_{{P}\times(M+1-P)}\right] \Sbf^{\textrm{T}}_{} \; .
    \label{eq: Positions Reconstruction}
\end{equation}
Using \eqref{eq: Pbf rel}, the (absolute) source position vector of the $s$-th source $\pbf_{s}$ can be obtained from $\Pbf_{\textrm{r}s}$ as
\begin{equation}
    \pbf_{s} \; = \; \underbrace{ \R \Pbf_{\textrm{r}s} }_{\Pbf_{s}}  \overline{\e}_{M+1}^{} \; ,
    \label{eq: Source absolute position}
\end{equation}
\color{black}with $\overline{\e}_{M+1}^{}$ the $(M+1)$-dimensional selection vector, consisting of zeros except for a one in the $(M+1)$-th entry.\color{black}

%% file: Sec_EDM_based_Position_Estimation.tex
\section{Euclidean Distance Matrix-Based Position Estimation}
\label{sec: EDM-Based Position Estimation}
In Section \ref{sec: Single-Source Position Estimation}, we review the EDM-based method in \cite{brumann20223d} to estimate the 3D position of a single source. 
Considering multiple candidate TDOA estimates, this method estimates the distance between the source and a reference microphone by minimizing an EDM-based cost function in a single continuous variable, considering all possible combinations of candidate TDOA estimates. 
In Section \ref{sec: EDM Multi-Source Position}, we propose an extension of this method to estimate the 3D positions of multiple sources. 
The proposed method estimates the optimal set of candidate TDOA estimates and corresponding source distances to the reference microphone by considering the $S$ smallest values of the EDM-based cost function. 
For each source, the absolute position is then computed from the estimated relative position by solving an orthogonal Procrustes problem. 

\subsection{Single-Source Position Estimation}
\label{sec: Single-Source Position Estimation}
In this section, we consider a single-source scenario, i.e., $S=1$. 
The EDM-based method in \cite{brumann20223d} estimates the source position $\pbf_{1}$ by first estimating the vector of squared distances $\dbf_{1}$ between the source and all microphones. 
To this end, an EDM-based cost function using the eigenvalues of the Gram matrix of the source is defined and minimized. 
In Section \ref{sec: EDM-Based Cost Function} we define the cost function assuming the TDOAs to be known. 
Aiming at improving robustness against noise and reverberation, in Section \ref{sec: TDOA Candidate Selection} we explain how to incorporate multiple TDOA estimates. 

\subsubsection{EDM-Based Cost Function}
\label{sec: EDM-Based Cost Function}
By introducing the variable $\alpha$, which represents the distance between the source and the reference microphone, the distance between the source and the $m$-th microphone can be written using \eqref{eq: d_ms} as a function of the variable $\alpha$ as 
\begin{equation}
    d_{m1}(\alpha) \; = \; \alpha \; + \; \nu \, \tau_{m}(\pbf_{1}) \;\;\, ,
    \label{eq: d_MS(alpha)}
\end{equation}
assuming the TDOAs $\tau_{m}(\pbf_{1}), \; m=2,\dots,M$, to be known (see Fig. \ref{fig: Near field}). 
Using \eqref{eq: d_MS(alpha)}, we construct the vector of squared distances as $\dbf_{1}(\alpha) = \left[d^{2}_{11}(\alpha), d^{2}_{21}(\alpha), \dots, d^{2}_{M1}(\alpha)\right]^{\textrm{T}}$ and the EDM in \eqref{eq: EDM} as a function of $\alpha$, i.e., 
\begin{equation}
{\D}_{1}(\alpha) \; = \; 
\begin{bmatrix}\renewcommand{\arraystretch}{1.3}
\begin{array}{c|c}
  \D_{MM}^{} & \dbf_{1}^{}(\alpha) \\
  \hline
  \dbf_{1}^{\textrm{T}}(\alpha) & 0_{}^{}
\end{array}
\end{bmatrix} \; .
\label{eq: EDM alpha}
\end{equation}
The Gram matrix ${\G}_{1}(\alpha)$ corresponding to the EDM ${\D}_{1}(\alpha)$ in \eqref{eq: EDM alpha} can be constructed using \eqref{eq: EDM to Gram}. 
Based on the fact that the rank of the Gram matrix $\G_{1}(\alpha)$ attains its minimum value (at most $P$) when $\alpha = d_{11}$, and is larger for other values of $\alpha$, it was proposed in \cite{brumann20223d} to define a cost function based on all but the $P$ largest eigenvalues $\lambda_{i}(\alpha)$ of $\G_{1}(\alpha)$, i.e., 
\begin{equation}
\boxed{
J_{}\left(\alpha\right) \;\;\; = 
\sum_{i=P+1}^{M+1} 
 |\lambda_i\left(\alpha\right)| 
}
\label{eq: source-ref-distance estimation_}
\end{equation}
It should be noted that since it cannot be guaranteed that the eigenvalues of ${\G}_{1}(\alpha)$ are positive for all values of $\alpha$ (e.g., in case of a mismatch between the distance variable and the TDOAs), the absolute values of the eigenvalues are used in \eqref{eq: source-ref-distance estimation_}. 
If $\alpha = d_{11}$, all but the $P$ largest eigenvalues of ${\G}_{1}(d_{11})$ are equal to zero, such that $J_{}\left(d_{11}\right) = 0$. 
The optimal value $\alpha_{\textrm{opt}} = d_{11}$ can hence be found as 
\begin{equation}
\alpha_{\textrm{opt}} \; = \;\; 
\argmin_{\alpha} 
\;\; J_{}\left(\alpha\right) \; .
\label{eq: source-ref-distance estimation}
\end{equation}
It should be noted that no closed-form solution is available, such that an exhaustive search over the (single) continuous variable $\alpha$ needs to be performed. 

\subsubsection{TDOA Estimation and Selection}\label{sec: TDOA Candidate Selection}
Since in practice the TDOAs of the source are obviously not available, in \cite{brumann20223d} a method was proposed to incorporate TDOA estimates into the EDM-based cost function. 
A commonly used method to estimate TDOAs between microphone pairs is based on the generalized cross correlation with phase transform (GCC-PHAT) \cite{knapp1976generalized}. 
The continuous-time GCC-PHAT function between the $m$-th and the reference microphone is defined as 
\begin{equation}
    \xi_{m}(\tau) \; = \int_{-\omega_{0}}^{\omega_{0}} \psi_{m1}^{}(\omega) e^{-\ji \omega \tau} d\omega \; ,
    \label{eq: IFT of GCC-PHAT}
\end{equation}
with radial frequency $\omega$, $\ji = \sqrt{-1}$, and time lag $\tau$. 
The normalized phase spectrum $\psi_{m1}(\omega)$ in \eqref{eq: IFT of GCC-PHAT} is given by 
\begin{equation}
    \psi_{m1}^{}(\omega) \; = \; \frac{
    \EX\{ Y_{m}^{}(\omega) Y_{1}^{*}(\omega) \}
    }{
    | \EX\{ Y_{m}^{}(\omega) Y_{1}^{*}(\omega) \} |
    } \; ,
\label{eq: psi}
\end{equation}
where $Y_{m}(\omega)$ denotes the $m$-th microphone signal in the frequency-domain, $\{\cdot\}^{*}$ denotes the complex-conjugate operator, and $\EX\{\cdot{}\}$ denotes the expectation operator. 
The TDOA between the $m$-th microphone and the reference microphone is then estimated by determining the main peak of $\xi_{m}(\tau)$, i.e., 
\begin{equation}
    \hat{\tau}_{m} \; = \;\; \argmax_{\tau} \;\; \xi_{m}(\tau) \; .
    \label{eq: TDOA estimation}
\end{equation}
Although the PHAT weighting in \eqref{eq: psi} has been shown to improve robustness against reverberation and noise \cite{chen2006time, velasco2016proposal, zhang2008does}, acoustic reflections can introduce peaks in $\xi_{m}(\tau)$ that are higher than the peak corresponding to the direct source component\color{black}, which means that the direct path peak is not guaranteed to be the global maximum. 
\color{black}
Therefore, it was proposed in \cite{brumann20223d} to consider $C$ candidate TDOA estimates\footnote{It should be noted that a small number of candidate TDOA estimates, e.g., $C=2$, is typically sufficient.}, corresponding to the $C$ largest local maxima, \color{black} instead of considering only the global maximum of the GCC-PHAT function as in \eqref{eq: TDOA estimation}\color{black}, and let the EDM-based method determine the combination of candidate TDOA estimates which corresponds to the most feasible 3D source position\color{black}. 
For each microphone $m = 2, \dots, M$, the set of $C$ candidate TDOA estimates is denoted as $\hat{\mathcal{T}}_m = \{\hat{\tau}_m(1), \dots, \hat{\tau}_m(C)\}$, corresponding to the $C$ largest local maxima of the GCC-PHAT function $\xi_m(\tau)$. 
{No association between candidate TDOA estimates and the source is applied at the TDOA estimation stage. The EDM-based method associates the best combination of candidate TDOA estimates to the source by determining which of the $Q=C^{M-1}$ possible combinations of candidate TDOA estimates corresponds to the most feasible source and microphones geometry.} 
We define the combination vectors 
\begin{equation}
    \mathbf{c}(q) \; = \; \left[ \, c_{2}(q), \dots, c_{M}(q) \, \right] \;\;\; , \quad \; q = 1, \dots, Q \; ,
    \label{eq: C(q)}
\end{equation}
where the index $c_{m}(q) \in \{1, \dots, C\}$ refers to one of the $C$ largest local maxima of $\xi_m(\tau)$. 
\color{black} It should be noted that the number of combinations $Q$ increases exponentially with $M$, which can be computationally demanding if $M$ is large. \color{black}
For each possible combination of candidate TDOA estimates, the vector of squared distances between the source and the reference microphone is given by
\begin{equation}
    \dbf_{1}(\alpha,q) \; = \; \left[ d_{11}^{2}(\alpha,q), d_{21}^{2}(\alpha,q), \dots, d_{M1}^{2}(\alpha,q) \right]^{\textrm{T}} \;\; , \quad  q = 1, \dots, Q \; ,
    \label{eq: d(alpha,q)}
\end{equation}
where, similarly to \eqref{eq: d_MS(alpha)}, 
\begin{equation}
    d_{m1}(\alpha,q) \; = \; \alpha + \nu \hat{\tau}_{m}(c_{m}(q)) \;\; , \quad m=1,\dots,M \, , \; q=1,\dots,Q \; .
    \label{eq: d_{m}(alpha,q)}
\end{equation}
Using \eqref{eq: d(alpha,q)}, for each possible combination of candidate TDOA estimates we can construct the cost function in \eqref{eq: source-ref-distance estimation_} as
\begin{equation}
    J(\alpha,q) \;\;\; = \sum_{i=P+1}^{M+1} |\lambda_{i}(\alpha,q)| \;\; , \quad q = 1,\dots , Q \; ,
    \label{eq: J(alpha,q)}
\end{equation}
with $\lambda_{i}(\alpha,q)$ the eigenvalues of the Gram matrix $\G_{1}(\alpha, q)$ associated with the EDM 
\begin{equation}
{\D}_{1}(\alpha,q) \; = \; 
\begin{bmatrix}\renewcommand{\arraystretch}{1.3}
\begin{array}{c|c}
  \D_{MM}^{} & \dbf_{1}^{}(\alpha,q) \\
  \hline
  \dbf_{1}^{\textrm{T}}(\alpha,q) & 0_{}^{}
\end{array}
\end{bmatrix} \;\; , \quad q = 1,\dots , Q \; .
\label{eq: EDM alpha k}    
\end{equation}
\color{black}When the candidate TDOA estimates and distance variable $\alpha$ are inconsistent with a feasible 3D geometry of the source and the microphones, the Gram matrix deviates from its ideal structure, which manifests itself as nonzero eigenvalues $\lambda_{i}(\alpha, q)$, for $i > P$ and, in some cases, even as negative eigenvalues. 
Using absolute values in \eqref{eq: J(alpha,q)} ensures that the cost function penalizes the magnitudes of these deviations, regardless of whether they appear as positive or negative eigenvalues. In other words, the cost function \eqref{eq: J(alpha,q)} describes the degree of geometric inconsistency of the candidate solution.
\color{black}
The optimal combination of candidate TDOAs is then estimated by first determining the optimal distance variable for each combination, i.e.,
\begin{equation}
    \hat{\alpha}_{}(q) \; = \;\; \argmin_{\alpha} \;\; J(\alpha,q) \;\; , \quad q = 1, \dots , Q \; ,
    \label{eq: hat{alpha}(q)}
\end{equation}
and then determining the combination which results in the smallest value of the cost function at these solutions, i.e.,
\begin{equation}
    \hat{q}_{1} \; = \; \argmin_{q=1,\dots,Q} \; J_{}(\hat{\alpha}(q),q) \; ,
    \label{eq: hat{alpha}_{1}}
\end{equation}
with corresponding Gram matrix $\hat{\G}_{1} = {\G}_{1}(\hat{\alpha}(\hat{q}_1),\hat{q}_{1})$. 
It should be noted that due to possible TDOA estimation errors, $J_{}(\hat{\alpha}(\hat{q}_1),\hat{q}_{1})$ is not guaranteed to be $0$ unlike $J_{}({\alpha}_{\textrm{opt}})$ in \eqref{eq: source-ref-distance estimation}.

As explained in Section \ref{sec: Theory}, from the eigenvalue decomposition of the estimated Gram matrix $\hat{\G}_{1} = \hat{\Sbf}\hat{\Lambd}\hat{\Sbf}^{\textrm{T}}$, the relative microphones and source positions matrix can be estimated as in \eqref{eq: Positions Reconstruction} using the $P$ largest eigenvalues, i.e., 
\begin{equation}
    \hat{\Pbf}_{\textrm{r}1} \; = \; \left[ \textrm{diag}\left( \sqrt{\hat{\lambda}_{1}} , \dots, \sqrt{\hat{\lambda}_{P}} \right) \, ,\; \0_{{P}\times(M+1-P)} \right] \hat{\Sbf}^{\textrm{T}} \; .
    \label{eq: hat{Pbf}_{r}}
\end{equation}

It should be noted that due to estimation errors it cannot be guaranteed that the estimated relative microphone positions matrix $\hat{\Mbf}_{\textrm{r}1}$ (first $M$ columns of $\hat{\Pbf}_{\textrm{r}1}$) can be perfectly mapped to the known absolute microphone position matrix $\Mbf$ by rotation and/or reflection as in \eqref{eq: Pbf rel}. 
As proposed in \cite{brumann20223d}, the mapping between the estimated relative microphone positions $\hat{\Mbf}_{\textrm{r}1}$ and the absolute microphone positions $\Mbf$ can be computed by solving an orthogonal Procrustes problem \cite{schonemann1966generalized,dokmanic2015euclidean}. 
First, the singular value decomposition (SVD) of $\hat{\Mbf}_{\textrm{r}1}^{} \Mbf^{\textrm{T}}_{}$ is computed as 
\begin{equation}
    \hat{\Mbf}_{\textrm{r}1}^{} \Mbf^{\textrm{T}}_{} \; = \; \hat{\U}_{1}^{} \hat{\Q}_{1}^{} \hat{\V}_{1}^{\textrm{T}} \; ,
    \label{eq: Procrustes SVD}
\end{equation}
where $\hat{\Q}_{1}^{}$ contains the singular values and $\hat{\U}_{1}^{}$ and $\hat{\V}_{1}^{}$ contain the left and right singular vectors, respectively. 
The orthogonal mapping matrix $\hat{\R}_{1}^{}$ in \eqref{eq: Pbf rel}, is then computed using the orthogonal Procrustes problem solution as 
\begin{equation}
    \hat{\R}_{1}^{} \;=\; \hat{\V}_{1}^{} \hat{\U}_{1}^{\textrm{T}} \; ,
    \label{eq: Procrustes solution}
\end{equation}
and the absolute source position is computed using \eqref{eq: Source absolute position} as 
\begin{equation}
    \hat{\pbf}_{1} \;=\; \hat{\R}_{1}^{}\hat{\Pbf}_{\textrm{r}1} \overline{\e}_{M+1}^{} \; .
    \label{eq: hat{pbf}_{1}}
\end{equation}
{No additional regularization beyond the eigenvalue truncation in \eqref{eq: hat{Pbf}_{r}} is applied for solving the orthogonal Procrustes problem.}

\subsection{Multi-Source Position Estimation}
\label{sec: EDM Multi-Source Position}
In this section, we propose an extension of the EDM-based method presented in the previous section, to estimate the position of multiple sources. 
For each source, we now introduce the variable $\alpha_{s}$, which represents the distance between the $s$-th source and the reference microphone. 
Similarly as in \eqref{eq: d_MS(alpha)}, the distance between the $s$-th source and the $m$-th microphone can be written as a function of the variable $\alpha_{s}$ as $d_{ms}(\alpha_{s}) = \alpha_{s} + \nu \tau_{m}(\pbf_{s}) \, , \,\;\; \; m = 2, \dots , M \, , \,\;\; \; s = 1 , \dots , S$.
Using $\dbf_{s}(\alpha_{s}) \;=\; \left[ d_{1s}^{2}(\alpha_{s}), d_{2s}^{2}(\alpha_{s}) , \dots , d_{Ms}^{2}(\alpha_{s}) \right]^{\textrm{T}}$, we can construct the vector of squared distances for the $s$-th source. 

It is indeed possible to construct an $(M+S)\times (M+S)$-dimensional EDM for all sources, similarly to \eqref{eq: EDM alpha}, i.e., 
\begin{align}\begin{split}
\D(\alpha_{1},\dots ,\alpha_{S},\D_{SS}) \;&= \\
& \hspace*{-1.2 cm}
\begin{bmatrix}\renewcommand{\arraystretch}{1.3}
\begin{array}{c|c}
\D_{MM} & 
\begin{array}{ccc}
\dbf_{1}(\alpha_{1}) & \cdots & \dbf_{S}(\alpha_{S}) 
\end{array}
\\\hline
\begin{array}{c} 
\dbf_{1}^{\text{T}}(\alpha_{1}) \\
\vdots \\
\dbf_{S}^{\text{T}}(\alpha_{S}) 
\end{array}
& 
\begin{array}{ccc}
\multicolumn{3}{c}{\D_{SS}} \\
\end{array}
\end{array}
\end{bmatrix} \; ,
\end{split}\end{align}
where it should be realized that this EDM depends on $S$ continuous distance variables and furthermore contains the $S \times S$-dimensional EDM $\D_{SS}$ with unknown inter-source distances. 
However, since jointly optimizing $\alpha_{1},\dots ,\alpha_{S}$ and the unknown entries of $\D_{SS}$ is not straightforward, we propose a simpler procedure, assuming that each source corresponds to a unique combination of TDOAs. 

{Similarly as for single-source position estimation, in the TDOA estimation stage, no explicit association between the candidate TDOA estimates and sources is performed.} 
We consider $C$ TDOA estimates for the microphones $m = 2, \dots, M${,} with $C \geq S$.  
For each combination of candidate TDOA estimates, the optimal value $\hat{\alpha}(q) \, , \; q = 1, \dots , Q, $ is determined using \eqref{eq: hat{alpha}(q)}. 
Instead of only considering the overall smallest value of the cost function at these solutions, as in the single-source case, 
{the candidate TDOA estimates are automatically associated to different sources by identifying the $S$ of $Q$ combinations with} the smallest values $J(\hat{\alpha}(\hat{q}_1),\hat{q}_{1}), \dots, \color{black}J(\hat{\alpha}(\hat{q}_S),\hat{q}_{S})\color{black}$ and their corresponding Gram matrices $\hat{\G}_{1}, \dots , \hat{\G}_{S}$. 
For each source, the relative microphones and source positions matrix $\hat{\Pbf}_{\textrm{r}s}$ is estimated based on the eigenvalue decomposition of the corresponding Gram matrix $\hat{\G}_{s}$, similarly to \eqref{eq: hat{Pbf}_{r}}. 
Then, similarly to \eqref{eq: hat{pbf}_{1}}, the position of the $s$-th source is estimated as 
\begin{equation}
    \hat{\pbf}_{\textrm{r}s} \;=\; \hat{\R}_{s}^{} \hat{\Pbf}_{\textrm{r}s}\overline{\e}_{M+1}^{} \; ,
    \label{eq: hat{pbf}_{textrm{r}s}}
\end{equation}
with $\hat{\R}_{s}$ the orthogonal mapping matrix for the $s$-th source, computed using the left and right singular vectors \color{black}of \color{black}$\hat{\Mbf}_{\textrm{r}s} \Mbf^{\textrm{T}}$. 

Since for reverberant and noisy scenarios with multiple sources the peaks of the GCC-PHAT functions may contain spurious peaks that don’t correspond to the true TDOAs, but to early reflections or noise, it may be beneficial to set $C>S$. 
\reb{This is especially true for spatially distributed microphones, where due to the potentially large distances between microphones, the range of plausible TDOAs in \eqref{eq: TDOA estimation} is larger than for compact microphone arrays.} The proposed EDM-based method can then be considered as a peak-to-source assignment method aiming at identifying the combinations of candidate TDOA estimates which correspond to the most feasible 3D geometry of sources and microphones. 
\reb{As a trade-off between performance and computational complexity, it will be shown in Sections \ref{sec: Computational Complexity Analysis} and \ref{sec: Position Performance Comparison} that setting $C=S+1$ is a good choice for multi-source position estimation.}
The EDM-based position estimation method is summarized in Algorithm \ref{tab: EDM Position Algorithm}. 
\begin{algorithm}[t]
\caption{\color{black}EDM-based multi-source position estimation}\color{black}
\label{tab: EDM Position Algorithm}
\begin{algorithmic}
\State \textbf{Parameters:} Number of microphones $M$, number of candidate TDOA estimates $C$, dimensionality $P$, number of sources $S$, combinations of candidate TDOA estimates $Q=C^{M-1}$.
\State \textbf{Input:} EDM of microphones $\D_{MM}$, vector of candidate TDOA estimate selection indices $\mathbf{c}(q)$, set of candidate TDOA estimates $\hat{\mathcal{T}}_m \; (m = 2, \dots, M)$, combination vectors $\cbf(q)$. 
\State \textbf{Output:} Estimated source positions $\hat{\pbf}_{1}, \dots, \hat{\pbf}_{S}$

\For{$q \; \leftarrow \; 1, \dots, Q$}
\For{$m \; \leftarrow \; 1, \dots, M$}
\State $\hat{d}_{m1}(\alpha,q) \; \leftarrow \; \alpha + \nu \hat{\tau}_{m}(c_{m}(q))$
\EndFor
\State $\hat{\D}_{}(\alpha,q) \; \leftarrow \; 
\begin{bmatrix}\renewcommand{\arraystretch}{1.3}
\begin{array}{c|c}
  \D_{MM}^{} & \hat{\dbf}_{}^{}(\alpha,q) \\
  \hline
  \hat{\dbf}_{}^{\textrm{T}}(\alpha,q) & 0_{}^{}
\end{array}
\end{bmatrix}$

\State $\hat{\G}_{}(\alpha,q) \; \leftarrow \; -\frac{1}{2} \, (\I_{M+1}-\frac{1}{M}\1_{M+1}^{}[\1_{M}^{\textrm{T}},0]) \, \hat{\D}_{}(\alpha,q) \, \dots $\\
$\quad\quad\quad\quad\quad\quad\quad \dots \; (\I_{M+1}^{}-\frac{1}{M}[\1_{M}^{\textrm{T}},0]^{\textrm{T}}\1_{M+1}^{\textrm{T}})$

\State $\hat{\Sbf}(\alpha,q)\hat{\Lambd}(\alpha,q)\hat{\Sbf}^{\textrm{T}}(\alpha,q) \; \leftarrow \; \hat{\G}_{}(\alpha,q)$

\State $J(\alpha,q) \;\;\;\leftarrow\; \sum_{i=P+1}^{M+1} |\hat{\lambda}_{i}(\alpha,q)|$


\State $\hat{\alpha}_{}(q) \;\leftarrow\;\; \argmin_{\alpha} \;\; J(\alpha,q)$

\EndFor

\State Determine the $S$ combinations $\hat{q}_1 , \dots , \hat{q}_S$ which result in the smallest cost function values $J(\hat{\alpha}_{}(\hat{q}_1),\hat{q}_1) , \dots, J(\hat{\alpha}_{}(\hat{q}_S),\hat{q}_S)$ 
\For{$s \; \leftarrow \; 1, \dots, S$}
\State $\hat{\Pbf}_{\textrm{r}s} \; \leftarrow \; \Bigl[ \textrm{diag}\Bigl( \sqrt{\hat{\lambda}_{1}(\hat{\alpha}_{}(\hat{q}_s),\hat{q}_s)} , \dots, \sqrt{\hat{\lambda}_{P}(\hat{\alpha}_{}(\hat{q}_s),\hat{q}_s)} \Bigr) \; \dots$ \\
$\quad\quad\quad\quad\quad \dots \; \0_{{P}\times(M+1-P)} \Bigr] \hat{\Sbf}^{\textrm{T}}(\hat{\alpha}_{}(\hat{q}_s),\hat{q}_s)$
\State $\hat{\Mbf}_{\textrm{r}s}^{} \;\leftarrow\; \hat{\Pbf}_{\textrm{r}s} [\I_{M} \;, \;\; \0_{M \times 1}]^{\textrm{T}}$
\State $\hat{\U}_{s}^{} \hat{\Q}_{s}^{} \hat{\V}_{s}^{\textrm{T}} \;\leftarrow\; \hat{\Mbf}_{\textrm{r}s}^{} \Mbf^{\textrm{T}}_{}$
\State $\hat{\R}_{s} \;\leftarrow\; \hat{\V}_{s}^{} \hat{\U}_{s}^{\textrm{T}}$
\State $\hat{\pbf}_{s} \;\leftarrow\; \hat{\R}_{s}^{}\hat{\Pbf}_{\textrm{r}s} \overline{\e}_{M+1}^{}$
\EndFor
\end{algorithmic}
\end{algorithm}

\begin{figure*}[t!]
\includegraphics[width=\linewidth]{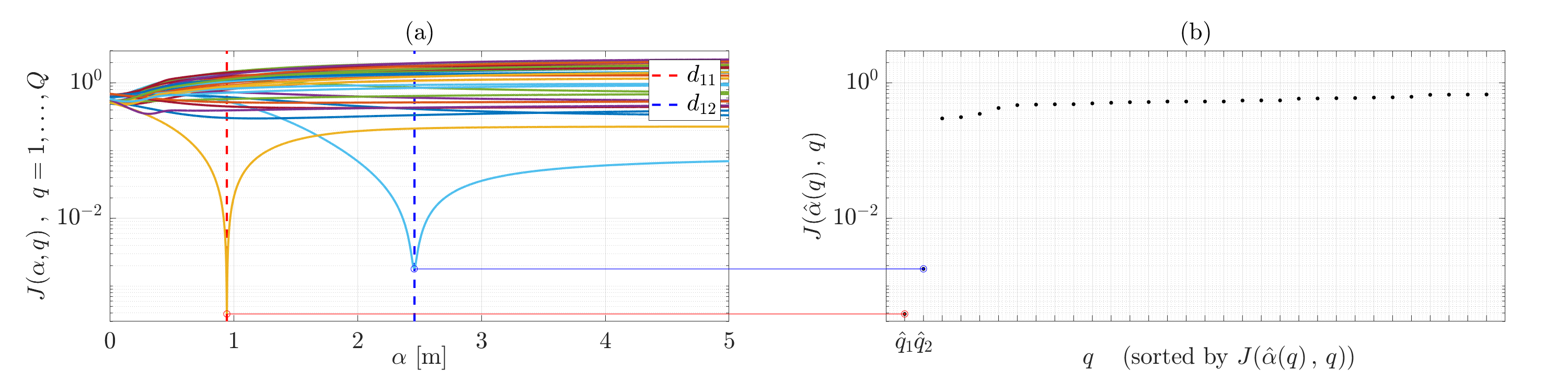}
\caption{(a) Cost functions $J(\alpha,q)$ for all combinations of TDOA estimates $q$, (b) Corresponding minimum cost function values.}
\label{fig: cost function multiple source pos}
\end{figure*}

For an exemplary scenario, Fig. \ref{fig: cost function multiple source pos}(a) illustrates the cost function $J(\alpha,q)$ in \eqref{eq: J(alpha,q)} for $M=6$ spatially distributed microphones and $S=2$ sources, while Fig. \ref{fig: cost function multiple source pos}(b) shows the sorted minimum cost function values $J(\hat{\alpha}(q),q)$. For this scenario, it can be observed that only 2 out of 32 cost functions exhibit clear minima. 
Moreover, it can be observed that the estimated distance variables $\hat{\alpha}_{1}$ and $\hat{\alpha}_{2}$ for these combinations of TDOA estimates correspond very closely to the distances $d_{11}$ and $d_{12}$.  

%% file: Sec_EDM_based_DOA_Estimation.tex
\section{Euclidean Distance Matrix-Based DOA Estimation}
\label{sec: EDM DOA Estimation}
Whereas in Section \ref{sec: EDM-Based Position Estimation} we considered spatially distributed microphones and proposed an EDM-based method to estimate the 3D positions of multiple sources, in this section we will consider compact microphone arrays and assume that the sources are in the far field of the microphone array. 
In Section \ref{sec: Single-Source DOA Estimation} we propose an EDM-based method to estimate the DOA of a single source. 
By defining a relative coordinate system in which one of the basis vectors is the DOA vector of the source, the rank of the Gram matrix associated with the EDM containing the microphone distances can be reduced by subtracting a rank-1 matrix based on the TDOAs. 
Using the eigenvalues of this rank-reduced Gram matrix, we define an EDM-based cost function to estimate the DOA, which does not require continuous variable optimization. 
In Section \ref{sec: Multi-Source DOA Estimation} we extend this method to estimate the DOAs of multiple sources. 
Similarly as for multi-source position estimation, we determine the optimal set of candidate TDOA estimates based on the EDM-based cost function, solving the association of candidate TDOA estimates to sources. 

\subsection{Single-Source DOA Estimation}
\label{sec: Single-Source DOA Estimation}
We consider a compact microphone array and a source in the far field of the microphone array, assuming that the acoustic waves arriving at the microphones from the source can be approximated as planar waves (see Fig. \ref{fig: Far field} for an exemplary two-dimensional configuration). 
This assumption is valid when the distances between the source and the microphones are much larger than the inter-microphone distances. 
In Section \ref{sec: Rank-Reduced Gram Matrix} we define a rank-reduced Gram matrix, assuming the TDOAs to be known. 
In Section \ref{sec: EDM DOA TDOA Estimation} we explain how to incorporate multiple TDOA estimates. 

\subsubsection{Rank-Reduced Gram Matrix}
\label{sec: Rank-Reduced Gram Matrix}
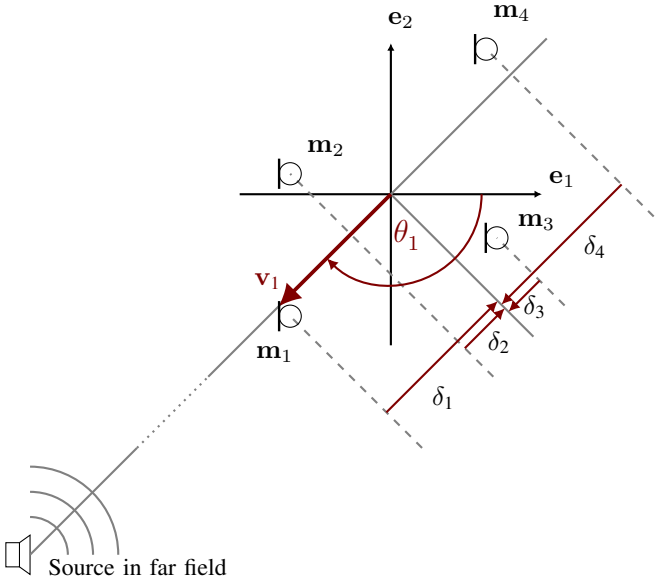
\begin{figure}
\centering
\begin{tikzpicture}[scale = 0.59]

\draw[black,thick,-{Latex[length=1mm, width=1mm]}] (20.125-3,-11.7) -- (20.125+3,-11.7); 
\draw[black,thick,-{Latex[length=1mm, width=1mm]}] (20.125,-11.7-3) -- (20.125,-11.7+3); 
\node[align=center] at (20.125+3+0.15,-11.7-0.45) {$\e_{1}$}; 
\node[align=center] at (20.125+0.6,-11.7+3-0.15) {$\e_{2}$};

\mic(17.1841,-12.7985)(0)(1);
\node[align=center] at (16.5439,-13.2933) {$\m_{1}$};
\mic(18.5983,-10.3491)(0)(1);
\node[align=center] at (19.4642,-10.2635) {$\m_{2}$};
\mic(21.5137,-13.5020)(0)(1);
\node[align=center] at (22.3516,-13.6183) {$\m_{3}$};
\mic(23.2038,-10.1503)(0)(1);
\node[align=center] at (23.9965,-9.7915) {$\m_{4}$};

\node[align=center] at (12.0732,-15.5961) {{\small{Source in far field}}};
\speaker(10.3449,-14.3206)(0)(0.6);
\centerarc[gray,thick](10.3449,-14.3206)(-30:60:0.75)
\centerarc[gray,thick](10.3449,-14.3206)(-30:60:1.25)
\centerarc[gray,thick](10.3449,-14.3206)(-30:60:1.75)

\draw[gray,thick,dashed] (18.1418,-16.3725) -- (17.1841,-12.7985); 
\draw[gray,thick,dashed] (24.4203,-14.6901) -- (23.2038,-10.1503); 
\draw[gray,thick,dashed] (22.0054,-15.3372) -- (21.5137,-13.5020); 
\draw[gray,thick,dashed] (20.0736,-15.8548) -- (18.5983,-10.3491); 
\draw[gray,thick,-] (21.1603,-15.5637) -- (20.1250,-11.7000); 

\draw[gray,thick,-] (10.3449,-14.3206) -- (13.2428,-13.5441);
\draw[gray,thick,dotted] (13.2428,-13.5441) -- (15.1747,-13.0264);
\draw[gray,thick,-] (15.1747,-13.0264) -- (24.3509,-10.5677);
\draw[red!50!black,line width=0.5 mm,{Latex[length=3mm, width=3mm]}-] (17.0807,-12.5157) -- (20.1250,-11.7000);
\node[red!50!black,align=center] at (17,-12.1) {$\vbf_{\textrm{1}}$};

\draw[red!50!black,thick,{Latex[length=1.5mm, width=1.5mm]}-] (20.9014,-14.5977) -- (17.8829,-15.4066);
\draw[red!50!black,thick,{Latex[length=1.5mm, width=1.5mm]}-] (20.9532,-14.7909) -- (19.8665,-15.0821);
\draw[red!50!black,thick,{Latex[length=1.5mm, width=1.5mm]}-] (20.9791,-14.8876) -- (21.8243,-14.6611);
\draw[red!50!black,thick,{Latex[length=1.5mm, width=1.5mm]}-] (20.9273,-14.6943) -- (24.1873,-13.8208);

\node[align=center] at (19.0042,-15.7273) {$\delta_{\textrm{1}}$};
\node[align=center] at (20.5134,-15.3229) {$\delta_{\textrm{2}}$};
\node[align=center] at (21.4793,-15.0641) {$\delta_{\textrm{3}}$};
\node[align=center] at (23.1093,-14.6274) {$\delta_{\textrm{4}}$};
\centerarc[red!50!black,thick,{Latex[length=1.5mm, width=1.5mm]}-](20.1250,-11.7000)(195:360:1.8)
\node[red!50!black,align=center] at (20.95,-12.5) {{\large{$\theta_{1}^{}$}}};

\end{tikzpicture}

\caption{Exemplary two-dimensional configuration, consisting of a compact array with $M=4$ microphones and a far-field source at azimuth angle $\theta_{1}^{}$. In the relative coordinate system, defined by the DOA vector $\vbf_{1}$, the relative distances between the $m$-th microphone and the centroid of the microphone array, in the direction of the source, are denoted by $\delta_{m}\, , \,\;\; m = 1, \dots, M$.
}
\label{fig: Far field}
\scriptsize
\end{figure}

In the absolute 3D coordinate system (with canonical basis vectors $\e_{1}^{}$, $\e_{2}^{}$, and $\e_{3}^{}$), the unit-norm DOA vector, pointing in the direction of the source (see Fig. \ref{fig: Far field}), 
is defined as 
\begin{equation}
    \vbf_{1}^{} \;=\; [ \textrm{cos}(\theta_{1}^{}) \textrm{cos}(\phi_{1}^{})  \; , \;\;\; \textrm{sin}(\theta_{1}^{}) \textrm{cos}(\phi_{1}^{}) \; , \;\;\; \textrm{sin}(\phi_{1}^{})]^{\textrm{T}} \; ,
    \label{eq: Cartesian DOA vector}
\end{equation}
with $\theta_{1}^{}$ and $\phi_{1}^{}$ the azimuth and elevation angle, respectively. 
These angles can be computed from the DOA vector as 
\noindent
\begin{minipage}{0.46\linewidth}
$$
\theta_{1}^{} \;=\; \textrm{tan}^{-1}\left( \frac{\e_{2}^{\textrm{T}}\vbf_{1}^{}}{\e_{1}^{\textrm{T}}\vbf_{1}^{}} \right) \; ,
$$
\end{minipage}
\hfill
\begin{minipage}{0.46\linewidth}
\begin{equation}
  \phi_{1}^{} \;=\; \sin^{-1}\left( \e_{3}^{\textrm{T}}\vbf_{1} \right) \; .
    \label{eq: Spherical coordinates 2}
\end{equation}
\end{minipage}\vspace{2mm}

We now define a relative coordinate system with orthonormal basis vectors $\e_{\textrm{x}}$, $\e_{\textrm{y}}$, and $\e_{\textrm{z}}$, which are not necessarily canonical, where we set one of the basis vectors equal to the DOA vector, e.g., $\e_{\textrm{x}} = \vbf_{1}$. 
The basis vectors of the relative coordinate system can be mapped to the basis vectors of the absolute coordinate system through a rotation with the angles $\theta_{1}^{}$ and $\phi_{1}^{}$, i.e., 
\begin{equation}
\e_{\textrm{x}} \;=\; \vbf_{1} \;=\; \R_{1}^{}\e_{1} \quad , \;\quad \e_{\textrm{y}} \;=\; \R_{1}^{}\e_{2} \quad , \;\quad \e_{\textrm{z}} \;=\; \R_{1}^{}\e_{3} \; ,\label{eq: e_{x}}
\end{equation}
with orthogonal mapping matrix 
\begin{equation}
\R_{1} = \begin{bmatrix}
\cos(\theta_{1}^{})\cos(\phi_{1}^{}) & -\sin(\theta_{1}^{}) & -\cos(\theta_{1}^{})\sin(\phi_{1}^{})\\
\sin(\theta_{1}^{})\cos(\phi_{1}^{}) & \cos(\theta_{1}^{}) & -\sin(\theta_{1}^{})\sin(\phi_{1}^{})\\
\sin(\phi_{1}^{}) & 0 & \cos(\phi_{1}^{}) 
\end{bmatrix} \; .
\end{equation}
The transpose of this matrix maps the microphone positions to the relative microphone positions as in \eqref{eq: Pbf rel}, i.e., $\Mbf_{\textrm{r}} = \R_{1}^{\textrm{T}} \Mbf$. 
Equivalently, the matrix $\Mbf_{\textrm{r}}$ describes the same microphone positions with respect to the absolute coordinate system as the matrix $\Mbf$ describes with respect to the relative coordinate system.
The $M$-dimensional relative coordinate vectors $\x_{\textrm{r}}$, $\y_{\textrm{r}}$, and $\z_{\textrm{r}}$ are defined as
\begin{equation}
    \Mbf_{\textrm{r}} \;=\; 
\begin{bmatrix}
     \x_{\textrm{r}}^{\textrm{T}} \\
     \y_{\textrm{r}}^{\textrm{T}} \\
     \z_{\textrm{r}}^{\textrm{T}}
\end{bmatrix} \; .
\label{eq: Mbf_{textrm{r}} xyz}
\end{equation}
It should be realized that the coordinate vector $\x_{\textrm{r}}$ contains the relative distances between the microphones and the centroid of the microphone array in the direction of the source  (see Fig. \ref{fig: Far field}). 
Since the relative distance $\delta_{m}$ between the $m$-th microphone and the centroid of the microphone array is directly related to the TDOA $\tilde{\tau}_{m} = \m_{m}^{\textrm{T}}\vbf_{1}/\nu$ between the $m$-th microphone and the centroid of the microphone array as $\delta_{m} = -\nu \tilde{\tau}_{m}$, the coordinate vector $\x_{\textrm{r}}$ can be written as 
\begin{equation}
\x_{\textrm{r}} \;=\; - \nu \tilde{\taubf} \; ,
\label{eq: x_{textrm{r}}}
\end{equation}
where $\tilde{\taubf} = \left[ \tilde{\tau}_{1}, \dots, \tilde{\tau}_{M} \right]^{\textrm{T}}$ denotes the vector of (centered) TDOAs, for which $\tilde{\taubf}^{\textrm{T}} \1_{M} = 0$. \color{black} As will be shown in the next subsection, these centered TDOAs can be estimated despite there not necessarily being a microphone at the centroid of the microphone array.\color{black} 

Using \eqref{eq: Mbf_{textrm{r}} xyz}, it can easily be seen that the $M \times M$-dimensional Gram matrix of the microphones ${\G}_{MM}$ in \eqref{eq: Gram rel} can be written as the sum of three rank-1 matrices, i.e., 
\begin{align}
    \G_{MM}^{} \; = \; \x_{\textrm{r}}\x_{\textrm{r}}^{\textrm{T}} + \y_{\textrm{r}}\y_{\textrm{r}}^{\textrm{T}} + \z_{\textrm{r}}\z_{\textrm{r}}^{\textrm{T}} \;, 
    \label{eq: Gram xyz}
\end{align}
whose rank is at most 3 (i.e., $P$). Assuming the TDOA vector $\tilde{\taubf}$ to be known, we now define the Gram matrix 
\begin{equation}
\boxed{
    \G_{MM}^{\;-} \;=\; \G_{MM} \;-\; \nu^2 \tilde{\taubf}\tilde{\taubf}^{\textrm{T}}
}
\label{eq: Rank-reduced Gram}
\end{equation}
Using \eqref{eq: x_{textrm{r}}} and \eqref{eq: Gram xyz}, it can be easily seen that $\G_{MM}^{\;-} = \y_{\textrm{r}}\y_{\textrm{r}}^{\textrm{T}} + \z_{\textrm{r}}\z_{\textrm{r}}^{\textrm{T}}$, whose rank is at most 2 (i.e., $P-1$). 
Considering the eigenvalue decomposition $\G_{MM}^{\;-} = \W^{} \Sigmabf^{} \W^{\textrm{T}}$
with the matrix $\W^{}$ containing the eigenvectors and the diagonal matrix $\Sigmabf^{}$ containing the eigenvalues, this means that at most $2$ eigenvalues are positive, $\sigma_{1} \geq \sigma_{2} \geq 0$, while the other eigenvalues are equal to zero, $\sigma_{3} = \dots = \sigma_{M} = 0$. 
By defining 
\begin{equation}
    \Mbf_{\textrm{r}}^{\;-} \;=\; 
\begin{bmatrix}
     \y_{\textrm{r}}^{\textrm{T}} \\
     \z_{\textrm{r}}^{\textrm{T}} 
\end{bmatrix} \; ,
\label{eq: Mbf_{textrm{r}} xy}
\end{equation}
the rank-reduced Gram matrix can be written as \color{black} 
\begin{equation}
\G_{MM}^{\;-} = (\Mbf_{\textrm{r}}^{\;-})^{\textrm{T}} \Mbf_{\textrm{r}}^{\;-} \; .
\end{equation}
As explained in Section \ref{sec: Theory}, a Gram matrix is only defined up to an arbitrary orthogonal transformation. 
This means that the rank-reduced Gram matrix $\G_{MM}^{\; -}$ can also be written as $\G_{MM}^{\; -} = \left(\Mbf_{\textrm{ar}}^{-}\right)^{\textrm{T}}\left(\Mbf_{\textrm{ar}}^{-}\right)$, with $\Mbf_{\textrm{ar}}^{-} = \left(\R_{\textrm{ar}}^{-}\right)^{\textrm{T}} \Mbf_{\textrm{r}}^{-}$, and $\R_{\textrm{ar}}^{-}$ an arbitrary orthogonal matrix. 
Based on \eqref{eq: Positions Reconstruction}, the matrix $\Mbf_{\textrm{ar}}^{\;-}$ can be computed as \color{black}
\begin{align}
    \Mbf_{\textrm{ar}}^{-} \;=\; \left[ \diag \left(
    \sqrt{\sigma_{1}}, \sqrt{\sigma_{2}}\right) \, , \; \0_{2 \times (M-2)}
     \right] \W^{\textrm{T}} \; \color{black}.\color{black}
    \label{eq: Mbf_{textrm{ar}2}}
\end{align}
\color{black}Using \eqref{eq: x_{textrm{r}}} and \eqref{eq: Mbf_{textrm{r}} xy}, the relative microphone positions can be written as
\begin{equation}
\renewcommand\arraystretch{1.3}
\Mbf_{\textrm{r}}^{} \;=\; 
\begin{lbmatrix}{1}
    \x_{\textrm{r}}^{\textrm{T}}\\
    \hline
    \Mbf_{\textrm{r}}^{-}
\end{lbmatrix} 
\;=\; 
\underbrace{
\left[\begin{array}{c|c}
1 & \0_{2}^{\textrm{T}} \\
\hline
\0_{2} & \R_{\textrm{ar}}^{-}
\end{array}\right]
}_{\R_{\textrm{ar}}}
\underbrace{\begin{lbmatrix}{1}
    -\nu \tilde{\taubf}^{\textrm{T}}\\
    \hline
    \Mbf_{\textrm{ar}}^{-}
\end{lbmatrix}
}_{\Mbf_{\textrm{ar}}} \; .
\label{eq: Mbf_{textrm{r}} from Mbf_{textrm{ar}}}
\end{equation}
Therefore, using \eqref{eq: Pbf rel}, the absolute microphone positions can be written as
\begin{align}\renewcommand\arraystretch{1.3}\Mbf \;=\; \R_{1} \Mbf_{\textrm{r}} \;=\; 
\underbrace{\R_{1} \R_{\textrm{ar}}
}_{\R'}
\Mbf_{\textrm{ar}} \; ,
\label{eq: R'}
\end{align}
with $\R'$ an orthogonal matrix \color{black}mapping the matrix $\Mbf_{\textrm{ar}}$ to the absolute microphone positions matrix $\Mbf$, which can be determined by solving an orthogonal Procrustes problem. \color{black}  
Realizing that the canonical basis vector $\e_{1}$ of the absolute coordinate system is equal to the first column of the matrix $\R_{\textrm{ar}}$ defined in \eqref{eq: Mbf_{textrm{r}} from Mbf_{textrm{ar}}}, i.e., $\e_{1} = \R_{\textrm{ar}} \e_{1}$, the DOA vector $\vbf_{1}$ in \eqref{eq: e_{x}} can be written as
\begin{align}\renewcommand\arraystretch{1.3}
\boxed{
\vbf_1 \;=\; \R_{1} \R_{\textrm{ar}} \e_{1} \;=\; \R'^{}  \e_{1} \; 
}
\label{eq: v_1 from R'}
\end{align}
This means that the DOA vector is equal to the first column of the \color{black}mapping \color{black}matrix $\R'^{}$ \color{black} and does not depend on the arbitrary orthogonal matrix $\R^{-}_{\textrm{ar}}$\color{black}.

\subsubsection{TDOA Estimation and Selection}
\label{sec: EDM DOA TDOA Estimation}
Computing the matrices $\G_{MM}^{\;-}$ in \eqref{eq: Rank-reduced Gram} and $\Mbf_{\textrm{ar}}$ in \eqref{eq: Mbf_{textrm{r}} from Mbf_{textrm{ar}}} requires centered TDOAs $\tilde{\taubf}_{m}$, which are not available in practice. 
Similarly as in Section \ref{sec: TDOA Candidate Selection}, we consider $C$ candidate TDOA estimates $\hat{\tilde{\taubf}}_{m} \; , $\linebreak{}$ m = 2, \dots, M$, corresponding to the $C$ largest local maxima of the GCC-PHAT function, and consider the $Q = C^{M-1}$ combination vectors $\cbf(q)$ in \eqref{eq: C(q)}. 
\color{black} Similarly as for position estimation, it should be noted that the number of combinations $Q$ increases exponentially with $M$. \color{black} 
For each combination $q$ of candidate TDOA estimates, 
the centered TDOA between the $m$-th microphone and the centroid of the microphone array can be estimated as 
\begin{equation}
    \hat{\widetilde{\tau}}_{m}^{}(q) \;=\; \hat{\tau}_{m}^{}(c_m^{}(q)) - \frac{1}{M}\sum_{j=1}^{M}\hat{\tau}_{j}^{}(c_j^{}(q)) \; .
    \label{eq: centered estimated TDOAs}
\end{equation}
The selection of candidate TDOA estimates will be explained in the next section for multiple sources (which can obviously also be used for a single source).

\subsection{Multi-Source DOA Estimation}
\label{sec: Multi-Source DOA Estimation}
In this section, we present a method to estimate the DOAs of multiple sources based on the eigenvalues of the Gram matrix defined in \eqref{eq: Rank-reduced Gram}. 
Contrary to the EDM-based position estimation method in Section \ref{sec: EDM-Based Position Estimation}, which requires an exhaustive search over the continuous variable $\alpha$, the proposed EDM-based DOA estimation method does not require continuous variable optimization. 

Using the vector of centered TDOA estimates $\hat{\tilde{\taubf}}(q) = [ \hat{\widetilde{\tau}}_{1}^{}(q), \dots, \hat{\widetilde{\tau}}_{M}^{}(q) ]^{\textrm{T}}$ based on \eqref{eq: centered estimated TDOAs}, the Gram matrix in \eqref{eq: Rank-reduced Gram} is defined for each possible combination of candidate TDOA estimates as
\begin{equation}
\hat{\G}_{MM}^{\;-}(q) \;=\; {\G}_{MM} - \nu^2 \hat{\tilde{\taubf}}(q)\hat{\tilde{\taubf}}^{\textrm{T}}(q) \;\;\; , \quad q = 1, \dots, Q \; .
\label{eq: Steered Gram EVD}
\end{equation}
Similarly to \eqref{eq: J(alpha,q)}, we define a cost function based on all but the $P-1$ largest eigenvalues $\hat{\sigma}_{i}(q)$ of $\hat{\G}_{MM}^{\;-}(q)$, i.e., 
\begin{equation}
\boxed{ 
I\left( q \right) \;\;= 
\sum_{i=P}^{M} 
 |\hat{\sigma}_{i}( q )| 
 \;  \;\;\; , \quad q = 1, \dots, Q 
}
\label{eq: DOA Estimation}
\end{equation}
For a single source ($S=1$), the optimal combination of candidate TDOA estimates is determined as the one minimizing the cost function, i.e.,  
\begin{equation}
\hat{q}_{1} \; = \;\; 
\argmin_{q=1,\dots,Q} 
\;\; I\left( q \right) \; .
\label{eq: centered-TDOA estimation}
\end{equation}
It should be noted that due to possible TDOA estimation errors the rank of $\hat{\G}_{MM}^{\;-}(\hat{q}_{1})$ is not guaranteed to be equal to $P-1$. 
For multiple sources, we consider the $S$ smallest cost function values $I(\hat{q}_{1}), \dots, I(\hat{q}_{S})$ and their corresponding Gram matrices $\color{black}\hat{\G}_{MM}^{\;-}(\hat{q}_{1}), \dots, \hat{\G}_{MM}^{\;-}(\hat{q}_{S})\color{black}$. Using the eigenvalue decomposition $\hat{\G}_{MM}^{\;-}(\hat{q}_{s}) = \hat{\W}(\hat{q}_{s})\hat{\Sigmabf}(\hat{q}_{s})\hat{\W}^{\textrm{T}}(\hat{q}_{s}) $, the matrices $\Mbf_{\textrm{ar}}^{-}$ in \eqref{eq: Mbf_{textrm{ar}2}} and $\Mbf_{\textrm{ar}}^{}$ in \eqref{eq: Mbf_{textrm{r}} from Mbf_{textrm{ar}}} can be estimated for each source as 
\begin{align}\begin{split}
    \hat{\Mbf}_{\textrm{ar}}^{-}(\hat{q}_{s}) \;&= \\
& \hspace*{-1.35 cm} \bigl[ \textrm{diag}\left( \sqrt{\hat{\sigma}_{1}(\hat{q}_{s})} , \dots, \sqrt{\hat{\sigma}_{P-1}(\hat{q}_{s})} \right) \, , \; \0_{{(P-1)}\times(M-P+1)} \bigr] \, \hat{\W}^{\textrm{T}}(\hat{q}_{s}) ,
    \label{eq: hat{Mbf}_{ar}^{-}}
\end{split}\end{align} 
\begin{equation}\renewcommand\arraystretch{1.4}
\hat{\Mbf}_{\textrm{ar}}^{}(\hat{q}_{s}) \; = \; 
\begin{lbmatrix}{1}
    -\nu \hat{\tilde{\taubf}}^{\textrm{T}}(\hat{q}_{s})\\
    \hline
    \hat{\Mbf}_{\textrm{ar}}^{-}(\hat{q}_{s})
\end{lbmatrix}.
\label{eq: hat{Mbf}_{ar}}
\end{equation} 
The mapping matrix $\hat{\R}'(\hat{q}_{s})$ between the matrix $\hat{\Mbf}_{\textrm{ar}}(\hat{q}_{s})$ and the absolute microphone positions matrix $\Mbf$ can be computed by solving an orthogonal Procrustes problem, similarly to \eqref{eq: Procrustes SVD} and \eqref{eq: Procrustes solution}. 
From these mapping matrices, the DOA vectors can then be estimated using \eqref{eq: v_1 from R'} as $\hat{\vbf}_{s} = \hat{\R}'(\hat{q}_{s}) \; \e_{1}$. No additional regularization beyond the eigenvalue truncation in \eqref{eq: hat{Mbf}_{ar}^{-}} is applied for solving the orthogonal Procrustes problem. 
\reb{Experimental results in Section \ref{sec: DOA Performance Comparison} indicate that setting $C=S$ is sufficient for multi-source DOA estimation using compact microphone arrays and that larger values of $C$ bring no notable benefit. 
This can be explained by the fact that the range of plausible TDOAs in the GCC-PHAT function \eqref{eq: TDOA estimation} is} \reb{much smaller for compact microphone arrays than for spatially distributed microphones, resulting in a low probability of picking up additional peaks due to early reflections or noise.}
\begin{algorithm}[t]
\caption{\color{black}EDM-based multi-source DOA estimation}\color{black}
\label{tab: EDM DOA Algorithm}
\begin{algorithmic}
\State \textbf{Parameters:} Number of microphones $M$, number of candidate TDOA estimates $C$, dimensionality $P$, number of sources $S$, combinations of candidate TDOA estimates $Q=C^{M-1}$.
\State \textbf{Input:} EDM of microphones $\D_{MM}$, vector of candidate TDOA estimate selection indices $\mathbf{c}(q)$, set of candidate TDOA estimates $\hat{\mathcal{T}}_m \; (m = 2, \dots, M)$, combination vectors $\cbf(q)$. 
\State \textbf{Output:} Estimated source DOA vectors $\hat{\vbf}_{1}, \dots, \hat{\vbf}_{S}$
\State $\G_{MM} \;\leftarrow\; -\frac{1}{2} \, (\I_{M}-\frac{1}{M}\1_{M}^{}\1_{M}^{\textrm{T}}) \, {\D_{MM}} \, (\I_{M}^{}-\frac{1}{M}\1_{M}\1_{M}^{\textrm{T}})$
\For{$q \; \leftarrow \; 1, \dots, Q$}
\For{$m \; \leftarrow \; 1, \dots, M$}
\State $\hat{\widetilde{\tau}}_{m}^{}(q) \;\leftarrow\; \hat{\tau}_{m}^{}(c_m^{}(q)) - \frac{1}{M}\sum_{j=1}^{M}\hat{\tau}_{j}^{}(c_j^{}(q))$
\EndFor
\State $\hat{\G}_{MM}^{\;-}(q) \;\leftarrow\; {\G}_{MM} - \nu^2 \hat{\tilde{\taubf}}(q)\hat{\tilde{\taubf}}^{\textrm{T}}(q) $
\State $\hat{\W}^{}\left( q \right) \hat{\Sigmabf}^{}\left( q \right) \hat{\W}^{\textrm{T}}\left( q \right) \; \leftarrow \; \hat{\G}_{MM}^{\;-}\left( q \right)$
\State $I\left( q \right) \;\;\leftarrow \;
\sum_{i=P}^{M} 
 |\hat{\sigma}_{i}( q )| $
\EndFor
\State Determine the $S$ combinations $\hat{q}_1, \dots, \hat{q}_{S}$ which result in the smallest cost function values $I\left( \hat{q}_1 \right), \dots, I\left( \hat{q}_{S} \right)$
\For{$s \; \leftarrow \; 1, \dots, S$}
\State $\hat{\Mbf}_{\textrm{ar}}^{-} \;\leftarrow \Bigl[ \textrm{diag}\left( \sqrt{\hat{\sigma}_{1}(\hat{q}_{s})} , \dots, \sqrt{\hat{\sigma}_{P-1}(\hat{q}_{s})} \right) \, , \;\; \dots $\\
$\quad\quad\quad\quad\quad\quad\quad \dots \;\;\0_{{(P-1)}\times(M-P+1)} \Bigr] \, \hat{\W}^{\textrm{T}}(\hat{q}_{s})$
\State $\hat{\Mbf}_{\textrm{ar}}^{} \; \leftarrow \; 
\begin{lbmatrix}{1}
    -\nu \hat{\tilde{\taubf}}^{\textrm{T}}(\hat{q}_{s})\\
    \hline
    \hat{\Mbf}_{\textrm{ar}}^{-}
\end{lbmatrix}$
\State $\hat{\U}_{s}^{} \hat{\Q}_{s}^{} \hat{\V}_{s}^{\textrm{T}} \;\leftarrow\; \hat{\Mbf}_{\textrm{ar}}^{} \Mbf^{\textrm{T}}_{}$
\State $\hat{\R}' \;\leftarrow\; \hat{\V}_{s}^{} \hat{\U}_{s}^{\textrm{T}}$
\State $\hat{\vbf}_s \; \leftarrow \; \hat{\R}'  \e_{1}$
\EndFor
\end{algorithmic}
\end{algorithm}

\begin{figure}[t!]
\centering
\includegraphics[width=\linewidth]{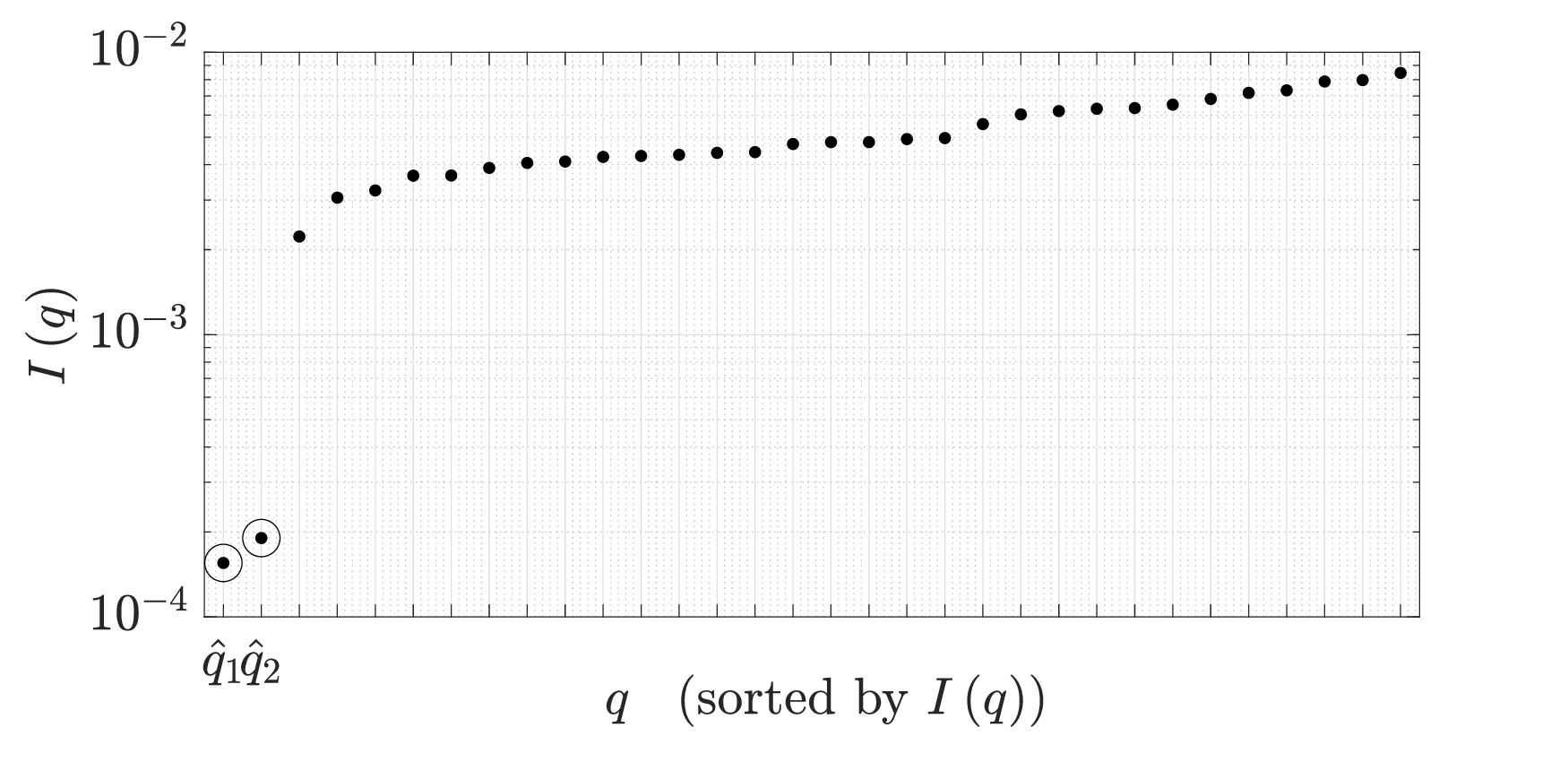}
\caption{Cost function values $I(q)$ for all combinations of candidate TDOA estimates $q$. 
}
\label{fig: DOA cost function}
\end{figure}

For an exemplary scenario, Fig. \ref{fig: DOA cost function} shows the sorted cost function values $I({q})$ for $M=6$ closely spaced microphones and $S=2$ sources. 
For this scenario, it can be observed that there is a clear difference between the two smallest values and the other values. 

%% file: Sec_SRP_PHAT.tex
\section{Baseline Localization Methods}\label{sec: SRP-PHAT}
In this section, we briefly review SRP-based methods to localize multiple sources, both for position estimation (Section \ref{sec: SRP-P}) and DOA estimation (Section \ref{sec: SRP-D}). 
These methods will be used as baselines in the experimental evaluation in the next section.
Contrary to the EDM-based methods proposed in Sections \ref{sec: EDM-Based Position Estimation} and \ref{sec: EDM DOA Estimation}, SRP-based methods require joint optimization of a functional over multiple continuous variables.  

\subsection{SRP-Based Multi-Source Position Estimation}\label{sec: SRP-P}
Similarly to the GCC-PHAT function in \eqref{eq: IFT of GCC-PHAT}, the SRP-PHAT functional for 3-dimensional position estimation \cite{grinstein2024steered} is defined as 
\begin{equation}
    \Psi(\pbf) \;\;= \sum_{i > j} \int_{-\omega_{0}}^{\omega_{0}} \psi_{ij}(\omega) e^{-\ji \omega \tau_{ij}(\pbf)} \, d\omega \; ,
\label{eq: SRP Integral}
\end{equation}
where $\tau_{ij}(\pbf)$ denotes the TDOA \eqref{eq: TDOA} corresponding to position $\pbf = [p_x, p_y, p_z]^{\textrm{T}}$, and the summation considers all microphone pairs, where $ i > j $. 
The estimated source positions $\hat{\pbf}_{1}, \dots, \hat{\pbf}_{S}$ are determined as the vectors that correspond to the $S$ largest local maxima of the SRP-PHAT functional. 
This requires a joint optimization of three continuous variables for all feasible positions within the room boundaries 
($ P_x^{\textrm{min}} \leq p_x \leq P_x^{\textrm{max}} $, $ P_y^{\textrm{min}} \leq p_y \leq P_y^{\textrm{max}} $, and $ P_z^{\textrm{min}} \leq p_z \leq P_z^{\textrm{max}} $). 
Practical considerations regarding the discretization of the continuous position variables and the determination of local maxima are discussed in Section \ref{sec: Implementation of SRP-Based Source Localization}. 

\subsection{SRP-Based Multi-Source DOA Estimation}\label{sec: SRP-D}
Similarly to the SRP-PHAT function for position estimation, 
the SRP-PHAT functional for DOA estimation is defined as 
\begin{equation}
    \Psi(\vbf) \;\;= \sum_{i > j} \int_{-\omega_{0}}^{\omega_{0}} \psi_{ij} (\omega) e^{-\ji \omega \tau_{ij}(\vbf)} \, d\omega \, ,
\label{eq: SRP Integral DOA}
\end{equation}
where $\tau_{ij}(\vbf) = (\m_{i} - \m_{j})^{\text{T}} \vbf / \nu$ denotes the TDOA corresponding to the DOA vector $\vbf$, which depends on the azimuth and elevation angles. 
The estimated azimuth and elevation angles $\hat{\theta}_{1}, \dots, \hat{\theta}_{S}$ and $\hat{\phi}_{1}, \dots, \hat{\phi}_{S}$ are determined from the DOA vectors that correspond to the $S$ largest local maxima of the SRP-PHAT functional. 
This requires a joint optimization of two continuous variables for all feasible azimuth angles $-\pi < \theta \leq \pi$ and elevation angles $-\pi/2 < \phi \leq \pi/2$. 
Practical considerations regarding the discretization of the azimuth and elevation angles and the determination of local maxima are discussed in Section \ref{sec: Implementation of SRP-Based Source Localization}.

%% file: Sec_Experimental_Evaluation.tex
\section{Experimental Evaluation}
\label{sec: Evaluation}
In this section, we compare the performance of the proposed EDM-based methods with the baseline SRP-based methods for multi-source position and DOA estimation in noisy and reverberant environments. 
We conducted two sets of experiments, where in experiment 1 we evaluated 3D position estimation using spatially distributed microphones and in experiment 2 we evaluated 3D DOA estimation using compact microphone arrays. 
Section \ref{sec: Acoustic Scenarios} describes the acoustic scenarios for both experiments. 
The implementation details of the EDM-based and SRP-based methods are presented in Sections \ref{sec: Implementation of EDM-Based Source Localization} and \ref{sec: Implementation of SRP-Based Source Localization}, respectively, and their computational complexity is compared in Section \ref{sec: Computational Complexity Analysis}.
The results of experiment 1 (position estimation) are discussed in Section \ref{sec: Position Performance Comparison}, while the results of experiment 2 (DOA estimation) are discussed in Section \ref{sec: DOA Performance Comparison}. 

\subsection{Acoustic Scenarios}
\label{sec: Acoustic Scenarios}
For both experiments, we considered a rectangular room with dimensions 6 m $\times$ 6 m $\times$ 2.4 m, $M=6$ microphones and $S=2$ static speech sources. 
For each experiment, 100 different acoustic scenarios were simulated, where the room impulse responses (RIRs) between the sources and the microphones were generated using the image source method \cite{allen1979image, HabetsRIR}, assuming equal reflection coefficients for all walls. 

\begin{figure*}
    \centering
    \hfill{}\begin{minipage}{0.43\linewidth}
        \includegraphics[width=\linewidth]{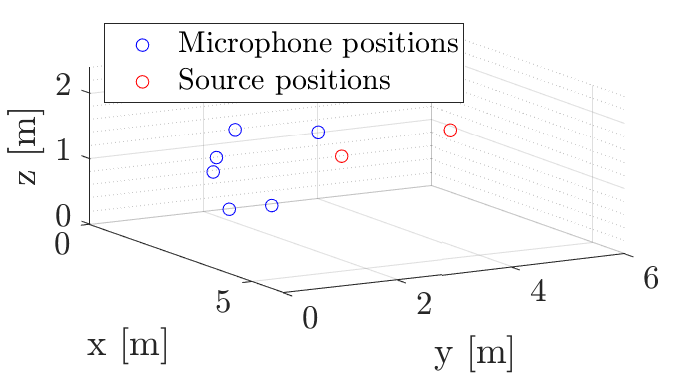}
    \end{minipage}\hfill{}\hfill{}
    \begin{minipage}{0.43\linewidth}
        \includegraphics[width=\linewidth]{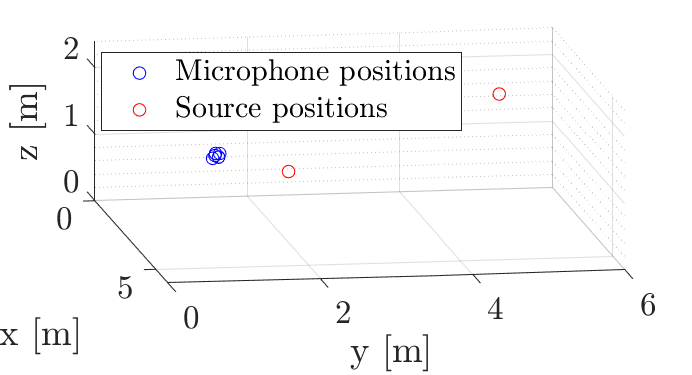}
    \end{minipage}\hfill{}
    \caption{
    {\color{black}Exemplary source and microphone constellations for experiment 1 (position estimation) and experiment 2 (DOA estimation), for $S=2$ sources and distance between the first source and centroid of the microphone array $d_{\textrm{c}1} = 4$ m.\color{black} }\color{black}\label{fig: Scenario}}
\end{figure*}

In experiment 1, the spatially distributed microphones were positioned randomly for each scenario within a cube with sides of length 2 m, with a minimum distance of 10 cm between microphones. 
In experiment 2, the microphones were positioned randomly for each scenario within a smaller cube with sides of length 10 cm, with a minimum distance of 4 cm between microphones. 
In both experiments, the distance between the second source and the centroid of the microphone array was fixed at $d_{\textrm{c}2} = 2 $ m. 
To evaluate the influence of source distance on localization accuracy, different distances $d_{\textrm{c}1}$ between the first source and the centroid of the microphone array were considered. 
In experiment 1, we considered $d_{\textrm{c}1} \in \{0, 1, 2, 3, 4\}$ m, where $d_{\textrm{c}1} = 0$ m and $d_{\textrm{c}1} = 1$ m\linebreak{} correspond to the first source being inside of the cube, while $d_{\textrm{c}1} = 3$ m and $d_{\textrm{c}1} = 4$ m correspond to the first source being outside of the cube. 
In experiment 2, we considered $d_{\textrm{c}1} \in \{0.5, 1, 2, 3, 4\}$ m, corresponding to both sources being outside of the cube (in the far field of the microphone array). 
For all scenarios, both sources were spaced at least 1 m apart and maintained a minimal angular separation of $20^{\circ}$ relative to the array centroid. 
Exemplary simulated scenarios for both experiments are plotted in Fig. \ref{fig: Scenario}. 

For each scenario, a 5-second speech signal, randomly selected from the M-AILABS dataset \cite{M-AILABS}, with equal probability of being a male or female speaker, was used for each source and convolved with the simulated RIRs. 
The sampling frequency was equal to $f_s = 16$ kHz. 
Spherically isotropic multi-talker babble noise generated using \cite{habets2008generating} was added to the reverberant speech mixtures at the microphones, with a reverberant signal-to-noise ratio (SNR) of 20 dB (averaged across the microphones). 
For each scenario, the reflection coefficients were set such that the direct-to-reverberant ratio (DRR) of the second source (at a fixed distance $d_{\textrm{c}2} = 2$ m) was equal to 5 dB (averaged across the microphones). 
The \color{black}resulting reverberation times were equal to $T_{60} \approx 186 \pm 16$ ms for experiment 1 and $T_{60} \approx 193 \pm 20$ ms for experiment 2. 
Obviously, the power ratio of the reverberant speech signals between the first and the second source (averaged across microphones and scenarios) decreased for increasing source distance $d_{\textrm{c}1}$, being approximately equal to 0 dB when $d_{\textrm{c}1} = d_{\textrm{c}2} = 2$ m.

\subsection{Implementation of EDM-Based Source Localization} 
\label{sec: Implementation of EDM-Based Source Localization} 

As explained in Sections \ref{sec: EDM-Based Position Estimation} and \ref{sec: EDM DOA Estimation}, the proposed EDM-based localization methods require candidate TDOA estimates, which are obtained from the GCC-PHAT function. 
In practice, the time-domain microphone signals are first transformed to the short-time Fourier transform (STFT) domain, with a frame length of $K=512$ samples (i.e., 32 ms), $50\%$ \color{black} overlap between frames, and using a square-root-Hann analysis window. 
Similarly to \eqref{eq: psi}, the instantaneous normalized phase spectrum between the $i$-th and $j$-th microphones is computed as
\begin{equation}
\psi_{ij}[k,l] \;=\; \frac{Y_{i}^{}[k,l] Y_{j}^{*}[k,l]}{|Y_{i}^{}[k,l] Y_{j}^{*}[k,l]|} \; , 
\label{eq: psi implemented}
\end{equation}
where $Y_{i}[k,l]$ denotes the STFT coefficient of the $i$-th microphone signal at frequency bin $k \in \{0, \dots, K-1\}$ and time frame $l \in \{1, \dots, L\}$, where $L$ denotes the number of frames {(i.e., $L=313$ frames for a 5-second signal)}. 
Similarly to \eqref{eq: IFT of GCC-PHAT}, the discrete-time GCC-PHAT function between the\linebreak{} $m$-th and the reference microphone is computed as 
\begin{equation}
    \xi_{m}^{}[n,l] \;\;= \sum_{k=0}^{K-1} \psi_{m1}^{}[k,l] e^{-\ji 2\pi n k / K} \;\;\; , \quad\; m = 2,\dots , M \;,
    \label{eq: IDFT of GCC-PHAT}
\end{equation}
where $n$ denotes the discrete-time index, with $\tau = n/f_s$. 
To achieve a more precise TDOA estimate, the discrete-time GCC-PHAT function $ \xi_{m}[n,l] $ is interpolated by a factor $ R = 20 $ using resampling. 
Only plausible time-lags $ n_m $ between the\linebreak[0]{} $ m $-th and the reference microphone are considered, determined by the inter-microphone distance $D_{m1}$, i.e., $ | n_m | <  R f_s D_{m1} / \nu $. 
To emphasize strong peaks, the function is weighted as $\tilde{\xi}_{m}[n_m,l] = \exp{(\gamma \xi_{m}[n_m,l])}$, with $\gamma = 30$ for position estimation (experiment 1) and $\gamma = 50$ for DOA estimation (experiment 2). 
Since the sources are assumed to be static, the GCC-PHAT function is averaged over all frames, i.e., 
\begin{equation}
    \tilde{\xi}_m[n_m] \;=\; \frac{1}{L} \sum_{l=1}^{L} \tilde{\xi}_{m}[n_m,l] \; .
    \label{eq: Batch GCC-PHAT}
\end{equation}
The $C$ discrete candidate TDOA estimates $\hat{n}_m^{}(1), \dots, \hat{n}_m^{}(C)$ between the $m$-th microphone and the reference microphone are determined from \eqref{eq: Batch GCC-PHAT} by using a peak-finding algorithm, which picks the $C$ highest local maxima. 
Each estimated TDOA is then fine-tuned using quadratic interpolation \cite{oppenheim2009discrete}. 
The continuous candidate TDOA estimates are then computed as $\hat{\tau}_{m}(c_m(q)) = \hat{n}_{m}(c_m(q))/(R f_s)$.

For EDM-based position estimation, the optimal distance variable in \eqref{eq: hat{alpha}(q)} was first determined using a one-dimensional grid search, with a resolution of 1 cm and a maximum distance of 6 m (i.e., {$G_{\alpha} = 601$ grid points}) and then fine-tuned with a quadratic interpolation. 
{$C\in \{2,3\}$ }candidate TDOA estimates were considered per microphone pair, resulting in {$Q = C^{M-1} \in \{32,243\}$} total combinations, respectively.
After computing the EDM-based cost functions $J(\alpha,q)$ in \eqref{eq: J(alpha,q)} and determining the optimal distance variable $\hat{\alpha}(q)$ using a one-dimensional grid search for all $Q$ possible combinations of candidate TDOA estimates, the respective cost function minima were sorted. 
The combination $\hat{q}_{1}$ yielding the smallest cost function minimum $J(\hat{\alpha}(\hat{q}_{1}),\hat{q}_{1})$ was used to estimate the source position $\hat{\pbf}_1$. 
To estimate the second source position $\hat{\pbf}_2$, only combinations where at least four candidate TDOA estimates differ were considered, as the likelihood that more than three TDOAs are shared between sources was assumed to be negligible. 
This rule prevents the EDM-based method from detecting sources that are spatially close. 

For EDM-based DOA estimation, {$C\in \{2,3\}$} candidate TDOA estimates were considered per microphone pair, resulting in {$Q\in \{32,243\}$} total combinations, respectively. 
After computing the EDM-based cost function values $I(q)$ in \eqref{eq: DOA Estimation} for all $Q$ possible combinations of candidate TDOA estimates, the respective cost function values were sorted. 
The combination $\hat{q}_{1}$ yielding the smallest value $I(\hat{q}_{1})$ was used to estimate the source DOA $\hat{\vbf}_1$. 
To estimate the second source DOA $\hat{\vbf}_2$, only combinations where at least four candidate TDOA estimates differ were considered.

It is known that the choice of reference microphone affects the TDOA estimation accuracy for single source scenarios \cite{brumann2025incremental}. 
Since preliminary experiments using multiple sources also demonstrated that the localization accuracy of the EDM-based methods may be negatively influenced by a poor choice of reference microphone, the reference microphone was chosen depending on the array geometry. 
For spatially distributed microphone arrays (experiment 1), the reference microphone was chosen as the microphone closest to the array centroid, reducing the likelihood that the reference microphone is located at a large distance from one or both sources which would result in a poor SNR. 
For compact microphone arrays (experiment 2), the reference microphone \color{black} for estimating the centered TDOAs $\hat{\tau}_{m}$ in \eqref{eq: centered estimated TDOAs} \color{black} was chosen as the microphone which was on average farthest from the other microphones, aiming at mitigating the influence of reverberation and noise on the reliability of the GCC-PHAT function \cite{brumann2024steered}. 

\subsection{Implementation of SRP-Based Source Localization} 
\label{sec: Implementation of SRP-Based Source Localization} 

For the SRP-based localization methods (Section \ref{sec: SRP-PHAT}), the exhaustive search over continuous variables is implemented as a discretized grid search. 
Using the instantaneous normalized phase spectrum in \eqref{eq: psi implemented}, the SRP-PHAT functionals for position and DOA estimation in Sections \ref{sec: SRP-P} and \ref{sec: SRP-D} are computed as 
\begin{equation}
\Psi[l]({\pbf}) \;\;\;\;\, = { \sum_{i > j}^{} \sum_{k=0}^{K-1} 
\psi_{ij} [k,l] e^{-\ji 2\pi f_s \tau_{ij}({\pbf}) k / K} } \;\; ,
\label{eq: SRP Integral practical}
\end{equation} 
\begin{equation}
\Psi[l]({\vbf}) \;\;\;\;\, = { \sum_{i > j} \sum_{k=0}^{K-1} 
\psi_{ij} [k,l] e^{-\ji 2\pi f_s \tau_{ij}({\vbf}) k / K} } \; .
\label{eq: SRP Integral practical DOA}
\end{equation} 
Similarly as for the GCC-PHAT function in \eqref{eq: Batch GCC-PHAT}, the SRP-PHAT functionals are averaged over all frames, i.e., 
\noindent
\begin{minipage}{0.48\linewidth}
$$
  \Psi(\pbf) = \frac{1}{L}\sum_{l=1}^{L} \Psi[l](\pbf) \; ,
$$
\end{minipage}
\hfill
\begin{minipage}{0.48\linewidth}
\begin{equation}
  \Psi(\vbf) = \frac{1}{L}\sum_{l=1}^{L} \Psi[l](\vbf) \; .
  \label{eq: SRP sum over frames DOA}
\end{equation}
\end{minipage}\vspace{2mm}
Exhaustively searching for the local maxima of these functionals can be computationally demanding if the grid resolution is high (especially for three-dimensional position estimation). 
Therefore, we first compute $\beta \geq S$ candidate positions/DOAs on a coarse grid (similarly to \cite{do2007fast}) that spans the entire range of feasible positions/DOAs. 
We then refine these coarse estimates by re-evaluating the functional on a finer grid in the vicinity of the coarse estimates. 
Finally, the $S$ positions $\hat{\pbf}_{1}^{\textrm{SRP}}, \dots, \hat{\pbf}_{S}^{\textrm{SRP}}$ or DOAs $\hat{\vbf}_{1}^{\textrm{SRP}}, \dots, \hat{\vbf}_{S}^{\textrm{SRP}}$ are estimated using the $S$ highest SRP-PHAT values on the finer grids. 

For position estimation, the functional in \eqref{eq: SRP sum over frames DOA} was first evaluated on a coarse grid with a 10 cm resolution along the x-, y-, and z-axes (i.e., $59 \times 59 \times 23 = 80,063$ grid points). 
The $ \beta = 3 $ coarse grid points with the highest SRP-PHAT values were then re-evaluated on a finer grid with a 1 cm resolution around those points, within a 20 cm $\times$ 20 cm $\times$ 20 cm cube (i.e., $3 \times 21^3 = 27,783$ grid points, {resulting in a total of $G_{\pbf} = 107,843$ grid points}). 
After estimating the first source position $\hat{\pbf}_{1}^{\textrm{SRP}}$ corresponding to the largest value of the fine grids, for the second source only coarse grid points at least 50 cm from the coarse estimate corresponding to the first source position were considered\color{black}. This rule prevents the SRP-based method from detecting sources that are spatially close\color{black}. 
For DOA estimation, the functional in \eqref{eq: SRP sum over frames DOA} was first evaluated on a coarse grid with a $ 5^{\circ} $ resolution in azimuth ($ \theta $) and elevation ($ \psi $) (i.e., $(72 \times 34) + 2 = 2,450$ grid points - noting that for elevation angles 0 and 180$^{\circ}$ the DOA is independent of the azimuth angle). 
The $ \beta = 2 $ coarse grid points with the highest SRP-PHAT values were then re-evaluated on a finer grid with a $ 0.5^{\circ} $ resolution, within $10^{\circ}$ in azimuth and elevation angles (i.e., $2 \times 21^2 = 882$ grid points, {resulting in a total of $G_{\alpha} = 3,332$ grid points}). 
After estimating the azimuth $\hat{\theta}_{1}^{\textrm{SRP}}$ and elevation $\hat{\psi}_{1}^{\textrm{SRP}}$ of the first source, for the second source only coarse grid points at least $ 20^{\circ} $ away from the coarse estimate corresponding to the first source DOA were considered (sometimes resulting in an additional coarse grid point being considered). 

\subsection{{Computational Complexity Analysis}} 
\label{sec: Computational Complexity Analysis}
{
This section analyzes the computational complexity of the EDM-based and SRP-based methods in terms of Big-O notation \cite{press2007numerical} and run time.
For the EDM-based method, the computational complexity associated with TDOA estimation is dominated by the computation of the resampled GCC-PHAT function \eqref{eq: IDFT of GCC-PHAT}, i.e., $\mathcal{O}(K R M \log (K R))$ per frame. 
The complexity of the subsequent EDM-based localization steps, using the candidate TDOA estimates based on the time-averaged GCC-PHAT function in \eqref{eq: Batch GCC-PHAT}, is dominated by the eigenvalue decomposition with complexity $\mathcal{O}(M^3)$, i.e., $\mathcal{O}\left( C^{M-1} M^3 G_{\alpha}\right)$ for position estimation and $\mathcal{O}\left( C^{M-1} M^3\right)$ for DOA estimation. 
The complexity of the SRP-based methods is given by 
$\mathcal{O}\left( G_{\pbf} M^2 L K \right)$ for position estimation and $\mathcal{O}\left( G_{\vbf} M^2 L K \right)$ for DOA estimation. 

It can be observed that the complexity of the EDM-based localization methods grows combinatorially with $C$ and $M$, which can become prohibitively large for a large number of microphones and/or when using many candidate TDOA estimates. 
In contrast, the complexity of the SRP-based methods only grows quadratically with $M$. 
In terms of discretized continuous variables, the complexity of the SRP-based methods depends on the resolution $G_{\pbf}$ of the three-dimensional variable $\pbf$ for position estimation and the resolution of $G_{\vbf}$ of the variable $\vbf$ for DOA estimation, which depends on the azimuth and elevation. 
In contrast, the complexity of the EDM-based position estimation method only depends on the resolution of the single variable $\alpha$ and the DOA estimation does not even require the discretization of a continuous variable. 

For the considered experimental setup with $M=6$ microphones and using the discretized grid values $G_{\alpha}$, $G_{\pbf}$, and $G_{\vbf}$ mentioned in the previous sections, we measured the run times of the EDM-based methods (\reb{for $C=2$ and $C=3$}) and the SRP-based methods using 5-second signals on a Ryzen 5900X processor. 
The average run times over all 100 simulated scenarios and 5 different distances $d_{\textrm{c}1}$ between source 1 and the array centroid are shown in Table \ref{tab: run times}. 
For the values of the variables defined in the previous subsections, the EDM-based position estimation method is around \reb{350} times faster than the SRP-based position estimation \reb{when using $C=2$, and around 280 times faster when using $C=3$}, and the EDM-based DOA estimation method is around \reb{25} times faster than the SRP-based method (\reb{for both $C=2$ and $C=3$}). 
It should be noted that in two-dimensional scenarios the number of grid points required for SRP-based position and DOA estimation is significantly smaller, which would correspondingly reduce the complexity and run time of the SRP-based approaches.
}


\begin{table}[t]
    \centering
    \caption{Average run times of the EDM-based and SRP-based position and DOA estimation methods.
    }
    \scriptsize
    \begin{tabular}{| c || c | c | c |} %
        \hline
         & \multicolumn{3}{c|}{Method run time [s] } \\\cline{2-4}
         & \reb{EDM$^{}$ ($C=2$)} & EDM$^{}$ ($C=3$) & SRP$^{}$ \\\hline\hline 
        Two-source position estimation & \reb{\textbf{1.6}} & \reb{\textbf{2.0}} & \reb{562.1}\\\hline 
        Two-source DOA estimation & \reb{\textbf{1.3}} & \reb{\textbf{1.3}} & \reb{32.7} \\\hline 
    \end{tabular} 
    \label{tab: run times}
\end{table} 
\subsection{Experiment 1: Position Estimation} 
\label{sec: Position Performance Comparison}
\begin{figure*}[!t]
\centering
\includegraphics[width=\linewidth]{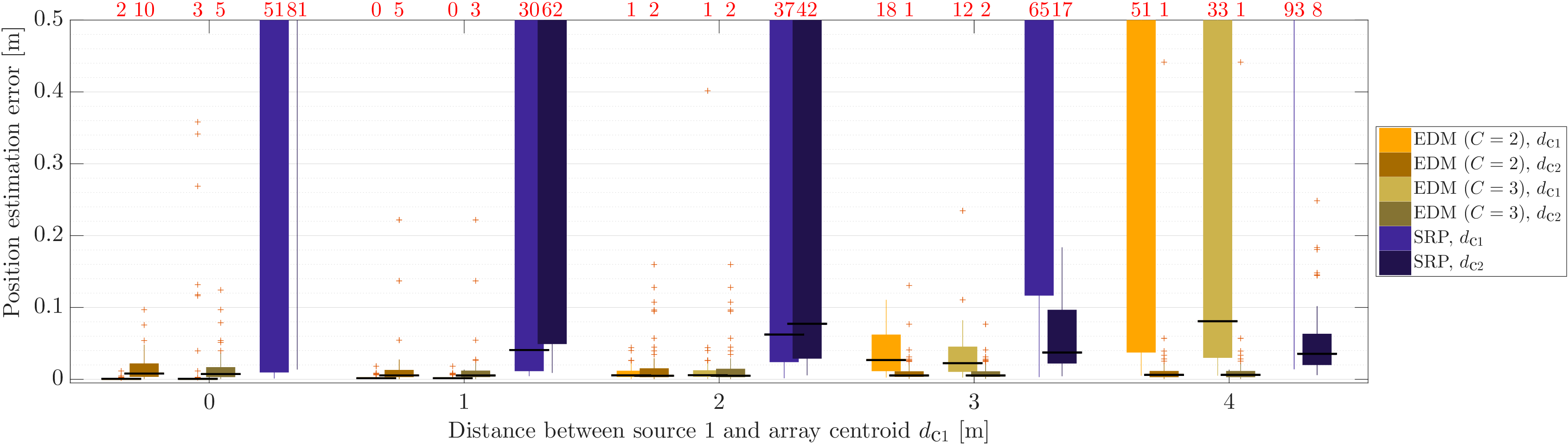}
\caption{\color{black}Box plots of position estimation errors for both sources. 
The position estimation errors are shown for the EDM-based method {(for $C \in\{2, 3\}$)} and SRP-based method, for different distances $d_{\textrm{c}1}$ between source 1 and the array centroid ($d_{\textrm{c}2} = 2$ m, 2 m cube).
The red numbers at the top denote the number of results outside of the plotted range.}
\label{fig: Position Results figure}
\end{figure*}

This section compares the position estimation accuracy of the proposed EDM-based method and the baseline SRP-based method for the acoustic scenarios with spatially distributed microphones described in Section \ref{sec: Acoustic Scenarios}. 
To evaluate the performance, we consider the position estimation error for each source, defined as $\varepsilon_{s}^{\textrm{pos}} = ||\pbf_{s} - \hat{\pbf}_{s}||_{2}^{} , \; s = 1,2,$
where the estimated source positions $\hat{\pbf}_{1}$ or $\hat{\pbf}_{2}$ are assigned to true source positions ${\pbf}_{1}$ or ${\pbf}_{2}$ using greedy assignment \cite{romeijn2000class} (i.e., assigning each estimated source to the closest true source sequentially, in order of increasing $\varepsilon_{s}^{\textrm{pos}}$). 
Fig. \ref{fig: Position Results figure} presents box plots of the position estimation errors $\varepsilon_{1}^{\textrm{pos}}$ and $\varepsilon_{2}^{\textrm{pos}}$ over 100 simulated scenarios, considering different distances $d_{\textrm{c}1}$ between source 1 and the array centroid. 
Table \ref{tab: Position Results table} summarizes the median position estimation errors for both sources. 
\begin{table}[t] 
\centering
    \caption{Median position estimation errors for the EDM-based and SRP-based methods for different distances $d_{\textrm{c}1}$ between\linebreak{} source 1 and the array centroid.}
    \scriptsize
    \begin{tabular}{| c || c | c | c | c | c | c |} 
        \hline
        &  \multicolumn{6}{c|}{Median position estimation error [cm]} \\\cline{2-7}
        & \multicolumn{2}{c|}{{EDM} {($C=2$)}}
        & \multicolumn{2}{c|}{EDM ($C=3$) }
        & \multicolumn{2}{c|}{SRP} \\\hline
        $d_{\textrm{c}1}$ [m]
        & {$\overline{\varepsilon}_{1}^{\textrm{pos}}$} & {$\overline{\varepsilon}_{2}^{\textrm{pos}}$} & {$\overline{\varepsilon}_{1}^{\textrm{pos}}$} & {$\overline{\varepsilon}_{2}^{\textrm{pos}}$} & {$\overline{\varepsilon}_{1}^{\textrm{pos}}$} & {$\overline{\varepsilon}_{2}^{\textrm{pos}}$} \\\hline\hline
        $0$ & \reb{\textbf{0.1}} & \reb{0.8} & \reb{\textbf{0.1}} & \reb{\textbf{0.7}} & \reb{60.0} & \reb{196.9}  \\\hline 
        $1$ & \reb{\textbf{0.2}} & \reb{\textbf{0.5}} & \reb{\textbf{0.2}} & \reb{\textbf{0.5}} & \reb{4.1} & \reb{144.8} \\\hline 
        $2$ & \reb{\textbf{0.6}} & \reb{\textbf{0.5}} & \reb{\textbf{0.6}} & \reb{\textbf{0.5}} & \reb{6.2} & \reb{7.7} \\\hline 
        $3$ & \reb{2.7}          & \reb{\textbf{0.5}} & \reb{\textbf{2.3}} & \reb{\textbf{0.5}} & \reb{182.0} & \reb{3.7} \\\hline 
        $4$ & \reb{94.3}        & \reb{\textbf{0.6}} & \reb{\textbf{8.1}} & \reb{\textbf{0.6}} & \reb{339.8} & \reb{3.6} \\\hline 
    \end{tabular}
    \label{tab: Position Results table}
\end{table}

As can be observed from Table \ref{tab: Position Results table}, for all source distances $d_{\textrm{c}1}$ the proposed EDM-based method yields considerably lower median position estimation errors for both sources than the SRP-based method. 
In addition, it can be observed from Fig. \ref{fig: Position Results figure} that the spread of the box plots is much smaller for the EDM-based method {(for both $C=2$ and $C=3$)} than for the SRP-based method for all source configurations. 
Furthermore, it is interesting to note that both methods behave quite differently for different source configurations. 
As can be observed from Table \ref{tab: Position Results table}, the median position estimation error for the EDM-based method mainly depends on source \color{black}distance, i.e., increases for source 1 with increasing $d_{\textrm{c}1}$, whereas it \color{black}remains relatively constant for source 2 (at $d_{\textrm{c}2} = 2$ m). 
On the other hand, the median position estimation error for the SRP-based method depends both on the absolute distances between the sources and the array centroid as well as on the relative distance between the sources. 
For instance, it can be clearly observed that the median position estimation error is larger for the source that is farther away from the array centroid (e.g., for $d_{\textrm{c}1} = 3$ m, $\overline{\varepsilon}_{1}^{\textrm{pos}}(\s) = \reb{182}$ cm and $\overline{\varepsilon}_{2}^{\textrm{pos}}(\s) = 3.7$ cm). 
This can be explained by the peaks of the farther source in the SRP functional being overpowered by the peaks of the closer source. 
In addition, when both sources are close to the microphones, the median position estimation error for both sources is large (e.g., for $d_{\textrm{c}1} = 0$ m,\linebreak{} $\overline{\varepsilon}_{1}^{\textrm{pos}}(\s) = \reb{60}$ cm and $\overline{\varepsilon}_{2}^{\textrm{pos}}(\s) = \reb{196.9}$ cm). 
This corresponds to \cite{garcia2021analytical, huang2021robust}, where it was shown that the accuracy of SRP-based position estimation degrades for sources located close to the microphone array because the peaks in the SRP-PHAT functional become narrow. 
This would necessitate a high 3D grid resolution to detect these peaks, which is computationally impractical in most applications. 
In contrast, the EDM-based method only requires a one-dimensional grid search on the distance variable $\alpha$, which is computationally more efficient to optimize at a high resolution.  Comparing the EDM-based methods for different numbers of candidate TDOA estimates, it can be observed that considering $C=3$ candidate TDOA estimates can slightly improve the position estimation accuracy compared to $C=2$, especially when one of the sources is far from the microphones (e.g., for $d_{\textrm{c}1} = 4$ m, $\overline{\varepsilon}_{1}^{\textrm{pos}} = \reb{94.3}$ cm for $C=2$ vs $\overline{\varepsilon}_{1}^{\textrm{pos}} = \reb{8.1}$ cm for $C=3$). 
\reb{Considering that choosing $C=3$ compared to $C=2$ only results in a minor increase in run time, from 1.6 s to 2 s as shown in Table \ref{tab: run times}, while slightly improving the position estimation performance, based on the results in Fig. \ref{fig: Position Results figure} and Table \ref{tab: Position Results table}, we recommend using $C=S+1$ for multi-source position estimation as a practical choice that balances position estimation performance and computational complexity.}
In conclusion, the consistently small median errors and small error distributions demonstrate the versatility of the proposed EDM-based multi-source position estimation method across a wide range of configurations. 

\color{black}\subsection{Experiment 2: DOA Estimation} 
\label{sec: DOA Performance Comparison}

\begin{figure*}[!t]
\centering
\includegraphics[width=\linewidth]{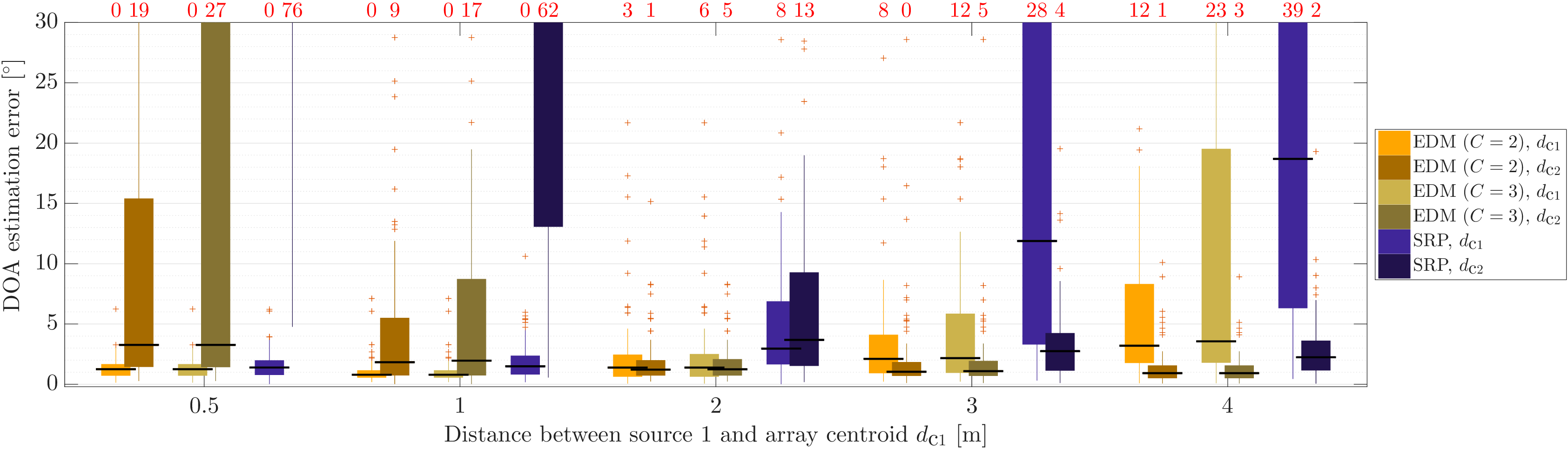}
\caption{Box plots of DOA estimation errors for both sources. 
The DOA estimation errors are shown for the EDM-based method {(for $C \in\{2, 3\}$)} and SRP-based method, for different distances $d_{\textrm{c}1}$ between source 1 and the array centroid ($d_{\textrm{c}2} = 2$ m, 10 cm cube). 
The red numbers at the top denote the number of results outside of the plotted range.}
\label{fig: DOA Results Figure}
\end{figure*}

This section compares the DOA estimation accuracy of the proposed EDM-based method and the baseline SRP-based method for the acoustic scenarios with compact microphone arrays described in Section \ref{sec: Acoustic Scenarios}. 
To evaluate the performance, we consider the DOA estimation error for each source, defined as $\varepsilon_{s}^{\textrm{DOA}} = \textrm{cos}^{-1}\left( 
    \hat{\vbf}_{s}^{\textrm{T}} \vbf_{s}^{}
    /
    (||\hat{\vbf}_{s}^{}||_{2}^{}\cdot||\vbf_{s}^{}||_{2}^{}
    ) \right) , \; s = 1,2 ,$ 
where the estimated DOA vectors $\hat{\vbf}_{1}^{}$ or $\hat{\vbf}_{2}^{}$ are assigned to true DOA vectors ${\vbf}_{1}^{}$ or ${\vbf}_{2}^{}$ using greedy assignment. 
Fig. \ref{fig: DOA Results Figure} presents box plots of the DOA estimation errors $\varepsilon_{1}^{\textrm{DOA}}$ and $\varepsilon_{2}^{\textrm{DOA}}$ over 100 simulated scenarios, considering different distances $d_{\textrm{c}1}$ between source 1 and the array centroid. 
Table \ref{tab: DOA Results table} summarizes the median DOA estimation errors for both sources. 
\begin{table}[t]
    \centering
    \caption{Median DOA estimation errors for the EDM-based and SRP-based methods for different distances $d_{\textrm{c}1}$ between\linebreak{} source 1 and the array centroid.}
    \scriptsize
    \begin{tabular}{| c || c | c | c | c | c | c |} 
        \hline
        &  \multicolumn{6}{c|}{Median DOA estimation error [$^{\circ}$]} \\\cline{2-7}
        & \multicolumn{2}{c|}{EDM ($C=2$)}
        & \multicolumn{2}{c|}{{EDM ($C=3$)}}
        & \multicolumn{2}{c|}{SRP} \\\hline
        $d_{\textrm{c}1}$ [m]
        & {$\overline{\varepsilon}_{1}^{\textrm{DOA}}$} & {$\overline{\varepsilon}_{2}^{\textrm{DOA}}$} & {$\overline{\varepsilon}_{1}^{\textrm{DOA}}$} & {$\overline{\varepsilon}_{2}^{\textrm{DOA}}$} & {$\overline{\varepsilon}_{1}^{\textrm{DOA}}$} & {$\overline{\varepsilon}_{2}^{\textrm{DOA}}$} \\\hline\hline 
        $0.5$ & \reb{\textbf{1.3}} & \reb{\textbf{3.3}} & \reb{\textbf{1.3}} & \reb{\textbf{3.3}} &  \reb{1.4} & \reb{50.2}  \\\hline 
        $1$   & \reb{\textbf{0.8}} & \reb{\textbf{1.8}} & \reb{\textbf{0.8}} & \reb{2.0} & \reb{1.5} & \reb{42.1} \\\hline 
        $2$   & \reb{\textbf{1.4}} & \reb{\textbf{1.2}} & \reb{\textbf{1.4}} & \reb{\textbf{1.2}} & \reb{3.0} & \reb{3.7} \\\hline 
        $3$   & \reb{\textbf{2.1}}          & \reb{\textbf{1.0}} &\reb{2.2} & \reb{1.1} & \reb{11.9} & \reb{2.8} \\\hline 
        $4$   & \reb{\textbf{3.2}} & \reb{\textbf{0.9}} & \reb{3.6}          & \reb{\textbf{0.9}} & \reb{18.7} & \reb{2.2} \\\hline 
    \end{tabular}
    \label{tab: DOA Results table}
\end{table}

As can be observed from Table \ref{tab: DOA Results table} for all source distances $d_{\textrm{c}1}$, the proposed EDM-based method {(for both $C=2$ and $C=3$)} yields median DOA estimation errors below $4^{\circ}$ for both sources, which are consistently smaller than the median DOA estimation errors obtained by the SRP-PHAT method. 
In addition, it can be observed from Fig. \ref{fig: DOA Results Figure} that the spread of the box plots is smaller for the EDM-based method than for the SRP-based method for all source configurations. 
It is also interesting to note that for both methods the median DOA estimation error is larger for the source that is farther away from the array centroid. 
This can be explained by the fact that the amplitudes of the peaks in the GCC-PHAT and SRP-PHAT functionals corresponding to both sources depend on their signal power ratio. 
This effect is much more pronounced for the SRP-based method than for the EDM-based method. 
For instance, when source 1 is close to the microphone array ($d_{\textrm{c}1} = 0.5$ m), the median DOA estimation error for source 1 is smaller than the median DOA estimation error for source 2 ($\reb{1.3} ^{\circ}$ and $\reb{3.3}^{\circ}$ for the EDM-based method {($C=2$)}; $\reb{1.4} ^{\circ}$ and $\reb{50.2}^{\circ}$ for the SRP-based method). 
In contrast, when source 1 is far from the microphone array ($d_{\textrm{c}1} = 4$ m), the median DOA estimation error for source 1 is larger than the median DOA estimation error for source 2 ($\reb{3.2} ^{\circ}$ and $0.9^{\circ}$ for the EDM-based method {($C=2$)}; $\reb{18.7} ^{\circ}$ and $\reb{2.2}^{\circ}$ for the SRP-based method). 
When both sources are equidistant ($d_{\textrm{c}1} = d_{\textrm{c}2} = 2$ m), both methods achieve similar median DOA estimation errors for both sources ($\reb{1.4} ^{\circ}$ and $\reb{1.2}^{\circ}$ for the EDM-based method {($C=2$)}; $\reb{3.0} ^{\circ}$ and $\reb{3.7}^{\circ}$ for the SRP-based method). 
For $d_{\textrm{c}1} = 0.5$ m, it can be seen that the median DOA estimation errors are also slightly larger than for $d_{\textrm{c}1} = 1$ m, reflecting that the far-field model assumption becomes less valid for small distances between the source and the microphones. 
Comparing the EDM-based methods for different numbers of candidate TDOA estimates, it can be observed that, unlike for spatially distributed microphones, considering $C=3$ candidate TDOA estimates does not improve DOA estimation accuracy compared to $C=2$. This is consistent with the fact that the number of spurious peaks in the GCC-PHAT functions is typically low for compact microphone arrays when only plausible time lags are considered. 
\reb{Since $C=S+1$ brings no notable performance benefit compared to choosing $C=S$, based on the results in Fig. \ref{fig: DOA Results Figure} and Table \ref{tab: DOA Results table}, we recommend choosing $C=S$ for multi-source DOA estimation.}

Similarly as for the position estimation results, the consistently small median DOA estimation errors and small error distributions demonstrate the versatility of the proposed EDM-based multi-source DOA estimation method across a wide range of source configurations. 